\documentclass[final,5p,times,twocolumn,authoryear]{elsarticle}

\journal{High Energy Astrophysics}

\usepackage{amssymb}
\usepackage{float}
\usepackage{lipsum}
\usepackage{ulem}
\usepackage{caption}
\usepackage{supertabular}

\usepackage{orcidlink}

\usepackage{longtable}
\usepackage{adjustbox}
\usepackage{rotating}
\usepackage{tikz}
\usepackage{geometry}
\usepackage{comment}

\usetikzlibrary{shapes.geometric, arrows, positioning}

\usepackage{graphicx}
\usepackage{tabularx}
\usepackage{natbib}
\usepackage{pdflscape}

\UseRawInputEncoding

\usepackage{amsmath}
\usepackage{hyperref}

\usepackage[T1]{fontenc}
\usepackage{subcaption}

\usepackage{makecell}
\usepackage{booktabs}
\usepackage{multirow}

\usepackage{color}

\def\inn#1{\textcolor{red}{[\textbf{IN}: #1]}}

\begin{document}
\begin{frontmatter}
\title{Creation of Viscous Dark Energy by the Hubble Flow: 
Comparison with SNe Ia Master Sample Binned Data}

\author[1]{Iolanda Navone\fnref{fn1}}
\fntext[fn1]{Corresponding author: navone.1867377@studenti.uniroma1.it}

\author[2,3,4]{Maria Giovanna Dainotti\corref{cor1}\fnref{fn2}}
\corref{cor1}
\fntext[fn2]{maria.dainotti@nao.ac.jp}

\author[1,5,6]{Elisa Fazzari\fnref{fn3}}
\fntext[fn3]{elisa.fazzari@uniroma1.it}

\author[1,7]{Giovanni Montani\fnref{fn4}}
\fntext[fn4]{giovanni.montani@enea.it}

\author[2,3]{Naoto Maki\fnref{fn5}}
\fntext[fn5]{naoto.maki@grad.nao.ac.jp}

\author[2,3]{Kazunori Kohri\fnref{fn6}}
\fntext[fn6]{kazunori.kohri@nao.ac.jp}

\affiliation[1]{Physics Department, Sapienza University of Rome, P.le A. Moro 5, 00185 Roma, Italy}
\affiliation[5]{Istituto Nazionale di Fisica Nucleare (INFN), Sezione di Roma, P.le A. Moro 5, I-00185, Roma, Italy}
\affiliation[6]{Physics Department, Tor Vergata University of Rome, Via della Ricerca Scientifica 1, 00133 Roma, Italy}
\affiliation[7]{ENEA, Nuclear Department, C.R. Frascati, Via E. Fermi 45, 00044 Frascati, Italy}
\affiliation[2]{Division of Science, National Astronomical Observatory of Japan, 2-21-1 Osawa, Mitaka 181-8588, Tokyo, Japan}
\affiliation[3]{Graduate University for Advanced Studies (SOKENDAI), 2-21-1 Osawa, Mitaka, Tokyo 181-8588, Japan}
\affiliation[4]{Space Science Institute, 4765 Walnut St Ste B, Boulder, CO 80301, USA}

\begin{abstract}
We study a family of cosmological models featuring dynamical dark energy (DE), based on the idea that the creation of its constituents arises from the gravitational field of the expanding universe, whose non-equilibrium physics is described by a non-zero bulk viscosity coefficient. We consider the complete scenario, in which both matter creation and bulk viscosity are present, together with its two limiting cases, in which only one of the two effects is retained. Once each model is constrained by requiring its present-day deceleration parameter $q_0$ to match specific values, the complete scenario introduces up to two additional free parameters with respect to the $\Lambda$CDM model, one of which is the equation of state parameter $w$ of the created dark energy. We consider two choices for $q_0$: the value predicted by the $\Lambda$CDM model, and one obtained from a background analysis of \citet{Fazzari:2025lzd}. To perform the analysis, we construct the effective running Hubble constant, i.e. a theoretical function corresponding to the ratio between the Hubble parameter of our models and the $\Lambda$CDM expansion rate. The theoretical predictions for the effective running Hubble constant of the three models are tested against the Master binned sample of Type Ia Supernovae (SNe Ia), through a Markov Chain Monte Carlo procedure with up to four free parameters. The most important result emerging from this analysis is that, when using the cosmographic $q_0$, all three models exhibit a quintessence-to-phantom transition in the effective equation of state parameter of the dark energy; on the contrary, when using the $q_0$ coming from the $\Lambda$CDM limit, the transition cannot happen, and the effective equation of state parameter is entirely of phantom nature across the considered redshift range.
\end{abstract}


\end{frontmatter}

\section{Introduction}

\noindent The Hubble constant tension, i.e. the $\simeq 4 - 6\, \sigma$ (depending on the dataset) inconsistency between the measurements of the Hubble constant ($H_0$) by the SH0ES Collaboration on Type Ia Supernovae (SNe Ia) \citep{SH0ES} and the one coming from the Planck Satellite observations of the Cosmic Microwave Background (CMB) \citep{Planck2018}, has become a promising subject of research due to its potential for bringing to light substantially novel physics.

A redshift tomography of the Hubble constant has been performed for the first time in \cite{dainotti2021}, revealing the intriguing result that the Hubble constant seems to be evolving with a power-law behaviour such as $\mathcal{H}_{0}(z=0)/(1+z)^\alpha$, where $\alpha$ mimics the evolutionary trend and $\mathcal{H}_{0}(z=0)$ is the value of the Hubble constant at $z=0$. 

The idea of this trend stems from the fact that many variables in astrophysics show redshift dependence with this behaviour; for studies in the domain of Gamma Ray Bursts (GRBs), see \citet{grb2022PASJ...74.1095D,opticalDainotti_2022,10.1093/mnras/stt1516,Dainotti_2013, 10.1093/mnras/stv1229,2017A&A...600A..98D}, 
while for Quasars see \citet{dainotti2024newbinningmethodchoose, galaxies12010004, qua2023ApJ...950...45D, qua2022ApJ...931..106D}.
We have already shown in \citet{10.1093/mnras/stt1516} that when evolutionary parameters of the physical relations (e.g. Phillips relation for SNe Ia, Dainotti relation for GRBs etc.) are not accounted for, a mismatch in the cosmological parameters occurs, namely we have an underestimation or overestimation of the values of $\Omega_{m0}$ and $H_0$ of the order of 10\%. Once these effects are considered we can have a better understanding of the physics at play and of the degeneracies among the evolutionary parameters and the cosmological ones. 

Another important point that is often overlooked is to check the statistical assumption used for the likelihoods in SNe Ia samples \citep{DAINOTTI202430,Dainotti_2023}. These new statistical analyses have revealed their efficacy even when combining multiple samples from different probes (see \citet{Dainotti_2023}). 

For more discussions on the Hubble tension that include GRBs and Quasars, and discussions on their role as cosmological probes, see also \citet{2023MNRAS.518.2201D,grb2014ApJ...783..126P,grb2022PASJ...74.1095D,grb2024JCAP...08..015A,qua2023ApJS..264...46L}; to instead explore the possibility of using Gravitational Waves (GWs) to shed light on the Hubble constant, see \citet{su2025exploringjointobservationcsst,pan2025determininghubbleconstantcrosscorrelation,zhan2025hubbleconstantmeasurementqpes}. Moreover, alternative probes such as the 21 cm brightness temperature seem to be highly sensitive to the value of $H_0$ in inhomogeneous cosmologies \citep{Mukherjee_2025}.
Additional late-time anomalies, such as the reported tension in the cosmic dipole, may also point toward new physics related to dark energy evolution, and GWs could provide an independent test of such anisotropies \citep{chen2025measuringcosmicdipolegolden}.
For recent studies showing a variety of different approaches to alleviate or solve the Hubble tension, see \citet{jia2025hubbletensionresolveddesi,Yarahmadi:2025fml, Yashiki:2025loj, goswami2025constrainingfrlmgravity, carloni2025addressingh0tensionmatter, lee2025alleviatinghubbletensioncosmological,du2025modelindependentlateuniversemeasurementsh0, articleGurzadyan, lee2025geometricinterpretationredshiftevolution, qua2023ApJS..264...46L, desimone2024doubletcosmologicalmodelschallenge,desmond2025subtlestatisticsdistanceladder, leclair2025quantumvacuumenergyorigin}. 

Interestingly, Baryonic Acoustic Oscillations (BAO) measurements of the Dark Energy Spectroscopic Instrument (DESI) \citep{DESI:2025zgx, DESI:2024mwx}, when combined with complementary cosmological probes, indicate that a dark energy (DE) component with a time evolving equation of state (EoS), described through the Chevallier-Polarski-Linder (CPL) parametrization \citep{CPL1,CPL2}, actually provides a better fit to the data than the standard $\Lambda$CDM model. The result suggests a preference for dynamical DE over a cosmological constant scenario.
 
This observational evidence, together with other experimental findings of a redshift-dependent $H_0$ (see also \citet{kalita2025revealinglimitationstandardcosmological}), is raising interest in phenomenological and physically grounded generalizations of the $\Lambda$CDM model, in which the DE term evolves with the redshift according to a given parameterization and specific dynamics (see \citet{divalentino-Hubbletension, CQG, CosmoVerseNetwork:2025alb, giare_overviewDDE, giare_dynamical, wang2025universeexperienceadslandscape,Wang:2025xvi, Montani_carlevaro_dainotti, Montani_deangelis_dainotti, Montani:2025rcy, Fazzari:2025mww, giare_interacting, divalentino_interacting, Zhai:2025hfi, Kessler:2025kju,manoharan2025solveshubbletensionphenomenological, Dixit_2025}). 

Furthermore, in \citet{Giare_Robust_DDE} it was shown that an evolutionary DE with a phantom crossing is the preferred scenario considering DESI data, regardless of the parametrization. This result is also supported by a model-independent cosmographic approach \citep{Fazzari:2025lzd} and by non-parametric methods to reconstruct the DE EoS (\cite{DESI:2025fii, Gonzalez-Fuentes:2025lei, Nesseris:2012tt, Berti:2025phi}). Such phantom behaviour cannot be achieved by quintessence models, which are described by a canonical scalar field \citep{Li:2011DE, Tsujikawa:2010DE}.
Therefore, the observational preference for a phantom crossing strongly motivates alternative mechanisms or exotic components (see \citet{efstratiou2025addressingdesidr2phantomcrossing}). 

Finally, the possible connection between dynamical DE and the Hubble tension is particularly relevant, since both may, in principle, point towards the need for late-time modifications of the standard cosmological model. For a more detailed discussion of this connection, see \citet{CosmoVerseNetwork:2025alb, DiValentino:2026uua} and references therein. 

An interesting phenomenological perspective on the possible dynamics of the Universe is instead presented in \citet{Linder:2024rdj}. The main point is that, although the CPL parametrization appears to be favoured over $\Lambda$CDM by current data combinations and may also help to alleviate the neutrino-mass tension that would otherwise arise \citep{Giare:2025ath}, it tends to worsen the Hubble tension. An interesting scenario with modified gravity and neutrinos has been developed in \citet{DOnofrio:2025cuk}, showing that it remains consistent with current observations and preferred over the $\Lambda$CDM model.

More theories implementing an evolving cosmological constant but in the early universe context can be found in \citet{1998GrCo....4S..50D,DYMNIKOVA_2000,Dymnikova:2001jy}; other scenarios leading to a global time-varying EoS parameter are also shown in \citet{10.1093/mnras/211.2.277,1985SvAL...11..236D,1988SvA....32..127D}.

In this context, an interesting diagnostic tool to measure deviations of data from a $\Lambda$CDM model is the so-called \textit{effective running Hubble constant}, introduced in \citet{dainotti2021} (see also \citet{schiavone2024}), where it has been used to show that the SNe Ia binned Pantheon Sample 
is compatible with the decaying power-law mentioned above. This result was confirmed and extended in \citet{dainotti2022}, where a varying-density matter-critical parameter is allowed in the fit procedure within each independent bin, and BAO data are added (for a similar formulation, see also \citet{kazantzidis2020}). The possibility of reproducing the same power-law behaviour in terms of a physical scenario has been successfully explored in \citet{schiavone2023}, where a metric $f(R)$-gravity in the Jordan frame has been implemented, and in \citet{Montani:2025rcy}, where a phenomenological interaction between DE and dark matter has been postulated. 

Recently, to provide a new arena for testing the effectiveness of the running Hubble constant, the Master Sample was created in \citet{Dainotti:2025qxz} and binned. The binning procedure has been explained and motivated in \citet{DAINOTTI202430}, where it is shown that the Gaussian likelihoods are not always the best-fit distributions.  

The binned SNe Ia data samples have been previously compared to dynamical DE models in \citet{Montani_carlevaro_dainotti, Montani_deangelis_dainotti} (for a metric $f(R)$-gravity formulation, see \citet{Montani_deangelis_fR}). In particular, in \citet{Fazzari:2025mww} it has been shown how the effective running Hubble constant is not only a useful diagnostic tool to detect the Hubble Tension, but also particularly efficient when investigating the phantom or quintessence nature of DE; in that analysis, using a DE model with matter creation in the $\Lambda$CDM limit, it has been concluded that the binned data of the SNe Ia Master Sample give no evidence of the phantom transition outlined by the DESI Collaboration for the chosen model (see also \citet{colgain_Dainotti}), nor show a preference for the dynamical DE scenario over the $\Lambda$CDM model. 

In the present paper, we develop a dynamical DE model in which both the matter creation \citep{matcre_calvaoLima, matcre_montani2001, elizalde_odintsov, Schiavone:2026agq, Fazzari:2025mww} and the bulk viscosity effects \citep{Brevik_2017,Brevik_2011,Belinskii1975,Capozziello_2006,Nojiri_2005,Belinskii1977,Belinskii1979, Montani_2017, Montani_carlevaro_dainotti,Carlevaro_2008} are taken into account for the dynamical picture. The motivation to simultaneously consider these two effects lies in the idea that the DE contribution is affected by a process of constituent creation driven by the gravitational field of the expanding universe, at a rate higher than $H_0$. As a result, this scenario could be characterized by a certain degree of non-equilibrium in thermodynamic phenomena, well described, as it is perturbative, by a bulk viscosity coefficient for the dark energy component \citep{Belinskii1979, Disconzi_2015}. The emerging effective running Hubble constant will be tested on the binned data of the SNe Ia Master Sample using the new cosmological model to test the nature of dark energy.

According to the analysis in \citet{Fazzari:2025mww}, where matter creation is present alone, the nature of the created DE has a slightly phantom character when imposing the $\Lambda$CDM limit. This differs from the case where bulk viscosity is taken as the only present dissipation under the same conditions: in fact, in this paper as well as in \citet{Montani_carlevaro_dainotti} (with a different viscosity coefficient), bulk viscosity alone seems to imply a quintessence nature for DE. To verify if this result holds generally, we will investigate what happens to the DE EoS when not imposing the $\Lambda$CDM limit, considering both the inclusion of bulk viscosity and matter creation into the dynamics.

The paper is structured as follows. In section \ref{sec:model}, we present the theoretical formulation of our dynamical dark energy models, also deriving the constraints imposed by requiring the deceleration parameter to match a specific value. The simplified cases with matter creation only and bulk viscosity only are examined as well, since they will be used for comparison. Finally, we introduce the effective running Hubble constant. Section \ref{sec:analysis} is devoted to the data analysis part, where we test the proposed scenario providing also a statistical comparison between different analysed models with different constraints, explaining the results. In section \ref{sec:conclusion} we summarize and further discuss the implications of our findings.

\section{\label{sec:model}Theoretical formulation}

\noindent We consider a flat, homogeneous and isotropic universe \citep{efstathiou_planck}, whose line 
element reads

\begin{equation}
	\text{d}s^2 = -\text{d}t^2 + a^2(t)\text{d}l^2
	\,, 
	\label{ed21}
\end{equation}
\noindent where $t$ denotes the synchronous time, $\text{d}l^2$ is the Euclidean infinitesimal distance element and $a(t)$ is the cosmic
scale factor, regulating the expansion of the universe and set equal to unity today.
The model studied in this paper considers the expansion as being driven by a pressureless matter energy density $\rho_m$ (including cold dark matter and baryonic matter), and by a dynamical dark energy contribution $\rho_{de}$. The energy density of radiation is neglected since this model will not be investigated outside the late universe. Hence, the Friedmann equation is stated as follows: 

\begin{equation}
	H^2 \equiv \left( \frac{\dot{a}}{a}\right)^2 = \frac{\chi}{3}\left( \rho_m + \rho_{de}\right)
	\,, 
	\label{ed22}
\end{equation}
\noindent where the dot refers to differentiation with respect to $t$ and $\chi\equiv8\pi G$ denotes the Einstein constant. 

In the proposed scenario, we retain the matter contribution $\rho_m$ in its standard form, i.e. governed by the dynamics

\begin{equation}
	\dot{\rho}_m + 3H\rho_m=0 \, \Rightarrow \rho_m(z) = 
	\rho_{m0} (1+z)^3
	\,, 
\end{equation}
\noindent with $\rho_{m0}$ being the present-day value of the matter energy density and the redshift variable being $z\equiv 1/a - 1$. 
The dark energy contribution, on the other hand, is modeled on two assumptions: firstly, that dark energy can be created by the time-varying gravitational field of the expanding universe, and secondly, that the non-equilibrium physics governing this process can be accounted for by a bulk viscosity effect. The modified continuity equation for the dark energy density therefore reads: 

\begin{equation}
	\dot{\rho}_{de} = 
	- 3H\left[ (1+w)\left( 1 - \frac{\Gamma (H)}{3H}\right) \rho_{de}- 3H\xi(\rho_{de} )\right]
	\,, 
	\label{eq:rhopunto}
\end{equation}
\noindent where $\Gamma$ is the particle creation rate \citep{matcre_calvaoLima, matcre_montani2001, elizalde_odintsov, matcre_nunes2015}, while $\xi$ is the bulk viscosity coefficient and $w$ is the EoS parameter associated with $\rho_{de}$. If the created matter is actually dark energy, we expect $w< - 1/3$.
It is convenient to rewrite Eq. \eqref{eq:rhopunto} after defining $\Omega_{de}\equiv \chi \rho_{de}/3H_0^2$, where $H_0$ denotes $H(z=0)$:

\begin{equation}
	\dot{\Omega}_{de} = - 
	3H\left[ (1+w)\left( 1 - \frac{\Gamma (H)}{3H}\right)\Omega_{de} - 3H\xi (\Omega_{de})\right]
	\, .
	\label{ed25}
\end{equation}

\noindent In general, the two phenomenological functions $\Gamma (H)$ and $\xi (\Omega_{de})$ can be taken as power-laws of their own arguments \citep{Freaza:2002ic, Cardenas:2020grl, Schiavone:2026agq}. For this model, we will consider the following \textit{ansatz}: 

\begin{equation}
	\Gamma (H)=\Gamma_0=const.\,\,, 
	\quad 
	\xi (\Omega_{de}) = 
	\xi_0\Omega_{de}\,\,,\, \quad 
	\xi_0=const.
	\, 
	\label{ed27}
\end{equation}

The first assumption treats the matter creation rate as approximately constant during the late-time evolution, when the rate of the universe's expansion, which could influence the matter creation rate, does not evolve rapidly. Secondly, we assumed that bulk viscous effects associated with the presence of dark energy scale with its density, becoming increasingly relevant as the Universe approaches the dark-energy-dominated era. 

\noindent We can now rewrite Eq. \eqref{ed25} and construct the set of equations that fully describes the evolution of the universe: 

\begin{alignat}{2}
&\dot{\Omega}_{de}(z)=-3H\Bigg[(1+w)\left(1-\frac{H_{\Gamma}}{E(z)}\right)-H_{\xi}E(z)\Bigg]\,\Omega_{de}(z),
\label{eq:omegap}
\\[4pt]
&\Omega_{de}(z=0)=1-\Omega_{m0}\,,
\nonumber
\\[10pt]
&E^2(z)\equiv\left(\frac{H}{H_0}\right)^2=\Omega_{m0}(1+z)^3+\Omega_{de}(z)\, .
\label{eq:E2}
\end{alignat}
\noindent where $H_{\Gamma}\equiv \Gamma_0/3H_0$ and $H_{\xi}\equiv 3H_0\xi_0$ are two free dimensionless parameters, while $E$ denotes the universe expansion rate. This model will be referred to as $M[\xi,\Gamma]$.

\subsection{\label{sec:constraints}Constraints}
\noindent The model defined in Eq. \eqref{eq:E2} has five free parameters, $H_0$, $\Omega_{m0}$, $H_\Gamma$, $H_\xi$, $w$.
In the following, we use the present-day deceleration parameter to relate some of these quantities and constrain one of them as a derived parameter. The deceleration parameter \citep{Visser:2003vq, Visser:2004bf} is defined as

\begin{equation}
	q_0 = -1 +\frac{1}{2}(E^2)^{\prime}_{x=0}
	\,,
	\label{eq:q}
\end{equation}

\noindent where $x \equiv \ln(1+z)$ and $(...)^{\prime} \equiv \text{d}(...)/\text{d}x = - \dot{(...)}/H$. 
Once a value of $q_0$ is imposed, from Eqs. \eqref{eq:q}, \eqref{eq:E2}, \eqref{eq:omegap}, one obtains:

\begin{equation}
\begin{cases}
    A(q_0,\Omega_{m0})\equiv\frac{2(1+q_0)-3\Omega_{m0}}{3(1-\Omega_{m0})} \; \\
    H_{\xi}=(w+1)(1-H_{\Gamma})
    -A(q_0,\Omega_{m0}) \,.
    \end{cases}
\label{eq:H_xi_cosm}
\end{equation}
Therefore, once $q_0$ is specified, the model loses one independent degree of freedom: $H_\xi$ is no longer sampled as an independent parameter but is derived from $q_0$, $H_\Gamma$, $w$, and $\Omega_{m0}$.

We consider two different implementations of this constraint. In the first case, we adopt a cosmographic value of $q_0$ based on the analysis of \citet{Fazzari:2025lzd} and computed in \ref{app:background_preliminary} for each model. This case, denoted as $q_0=q_0^{\mathrm{cosm}}$, represents the more general choice, since the deceleration parameter is inferred from a model-independent late-time 
reconstruction rather than fixed to a value dictated by $\Lambda$CDM. In the second case, we assume that the model is sufficiently close to $\Lambda$CDM at low redshift, so that it shares the same present-day deceleration parameter,
\begin{equation}
q_0=q_0^{\Lambda\mathrm{CDM}}=-1+\frac{3}{2}\Omega_{m0}\,.
\label{eq:q0_LCDM}
\end{equation}

This prescription is adopted in order to ensure that the proposed model behaves well near $z\simeq0$, and, in particular, that it fulfils the luminosity-redshift relation \citep{Riess:2016jrr, Efstathiou:2021ocp}. It also allows us to compare our results with the constraints obtained in the previous work \citep{Fazzari:2025mww}. In this limit, the function $A(q_0,\Omega_{m0})$ vanishes, and Eq.~\eqref{eq:H_xi_cosm} reduces to

\begin{equation}	H_{\xi}\stackrel{\,\,\,\,\,\,\,\,\,q_0=q_0^{\Lambda \mathrm{CDM}}}{=}  (w+1)(1-H_{\Gamma}) .
	\label{eq:q0lcdmconstrain} 
\end{equation}
\noindent In other words, this condition forces our model to reproduce the $\Lambda$CDM behaviour at low redshift. We will therefore compare the results obtained under the cosmographic deceleration parameter, $q_0=q_0^{\mathrm{cosm}}$, with those obtained under the $\Lambda$CDM low-redshift condition, $q_0=q_0^{\Lambda\mathrm{CDM}}$.
In both cases, the effective EoS parameter $w_{\mathrm{eff}}(z)$ can be expressed as:
\begin{equation}
	w_{\mathrm{eff}}(z) = -1 
	+ \left[(1+w) \left( 1 - 
	\frac{H_\Gamma}{E(z)}\right) 
	- H_{\xi}E(z)\right]
	\, . 
	\label{eq:weff}
\end{equation}

Furthermore, our physical interpretation of $H_\Gamma$ motivated us to focus our search on the range $H_\Gamma > 1$. In fact, in order for the particle creation process to be efficient, it is reasonable to expect a creation rate ($\Gamma_0=3H_\Gamma{H_0}$) significantly higher than the expansion rate of the Universe (especially since the dark energy dominated phase of the Universe, in which this study takes place, only corresponds to the last four billion years).
\subsection{Isolated matter creation effect}\label{sec:nobulk}
\noindent To better understand the role of bulk viscosity, the model will be compared to the previously studied case \citep{Fazzari:2025mww}, here labeled $M[{\Gamma}]$, where only the matter creation process is considered.

Removing the bulk viscosity effect from Eq. \eqref{eq:omegap}, we get:
\begin{equation}
\dot{\Omega}_{de}(z)=-3H(1+w)\left(1-\frac{H_{\Gamma}}{E(z)}\right)\Omega_{de}(z)\,.
\label{eq:omegapmatt}
\end{equation}

\noindent By repeating the calculation in section \ref{sec:constraints}, we can easily express the matter creation parameter in terms of the deceleration parameter as:
\begin{equation}
    H_{\Gamma}=1-\frac{A(q_0,\Omega_{m0})}{w+1} \, \quad \text{with } \quad H_\xi=0 \,,
\end{equation}
\noindent and in the $\Lambda\mathrm{CDM}$ limit we get
\begin{equation}
  H_{\Gamma}\stackrel{\,\,\,\,\,\,\,\,\,q_0=q_0^{\Lambda \mathrm{CDM}}}{=}  1 \,.
  \label{eq:H_g_LCDM_1}
\end{equation}

\noindent Finally, the effective EoS parameter follows the equation:
\begin{equation}
w_{\mathrm{eff}}(z)=-1+(1+w)\left(1-\frac{H_\Gamma}{E(z)}\right) \,,
\label{eq:weffmatter}
\end{equation}
yielding
\begin{equation}
w_{\mathrm{eff}}(z)\stackrel{\,\,\,\,\,\,\,\,\,q_0=q_0^{\Lambda \mathrm{CDM}}}{=}-1+(1+w)\frac{E(z)-1}{E(z)}.
\label{eq:q0lcdmweffmatter}
\end{equation}
\subsection{\label{sec:onlybulk}Isolated bulk viscosity effect}
\noindent Let us now examine the opposite case, i.e. when bulk viscosity is included while matter creation is removed. This scenario is labelled as $M[\xi]$. 

Once again, constraining the deceleration parameter to a specific $q_0$ value,  Eq. \eqref{eq:H_xi_cosm} simplifies to 
\begin{equation}
H_\xi=1+w-A(q_0,  \Omega_{m0})\quad \text{with } \quad
H_\Gamma=0\,.
\end{equation}
\noindent Therefore, the differential equation to solve becomes (from Eq. \eqref{eq:omegap}):

\begin{equation}
	\dot{\Omega}_{de}(z)= -
	3H \big(1+w-H_\xi E(z)\big)\, \Omega_{de}(z)\,,
\end{equation}
\noindent with the effective EoS therefore being:

\begin{equation}
		w_{\mathrm{eff}}(z) = w  
	- H_{\xi}E(z)
	\, . 
    \label{eq:weffbulk}
\end{equation}

When imposing the $\Lambda$CDM limit, i.e. Eq. \eqref{eq:q0_LCDM}, one obtains:

\begin{equation}
H_\xi\stackrel{\,\,\,\,\,\,\,\,\,q_0=q_0^{\Lambda \mathrm{CDM}}}{=}1+w \,,
\label{eq:constraintwbulk}
\end{equation} 
and 

\begin{equation}
	w_{\mathrm{eff}}(z)\stackrel{\,\,\,\,\,\,\,\,\,q_0=q_0^{\Lambda \mathrm{CDM}}}{=}-1+(1+w) \,(1-E(z)).
    \label{eq:q0lcdmweffbulk}
\end{equation}

\subsection{\label{sec:level2}The effective running Hubble constant}

\noindent Following the example of previous studies 
(\cite{Fazzari:2025mww, Montani:2025rcy,PhysRevD.103.103509,krishnan2022h0universalflrwdiagnostic,schiavone2024}), we now introduce an object that has proven useful in highlighting the differences in behaviour between dynamical DE models and the $\Lambda$CDM model.
The \textit{effective running Hubble constant} is defined as:

\begin{equation}
\mathcal{H}_{0}(z) \equiv H_0\frac{E(z)}{E_{\Lambda \mathrm{CDM}}(z)}=H_0\frac{E(z)}{\sqrt{ \Omega_{m0} (1+z)^{3} + 1 - \Omega_{m0} }} \,,
\end{equation}

\noindent where $E_{\Lambda \text{CDM}}$ denotes the universe expansion rate according to the $\Lambda$CDM model in late universe dynamics. If the universe follows the $\Lambda$CDM model, this function is expected to reduce to the constant value $\mathcal{H}_{0}(z) \equiv H_0$. On the other hand, the appearance of a non-constant $\mathcal{H}_{0}(z)$ reflects the spread in the different experimental values of $H_0$ when observed in different redshift intervals.

For the dynamical dark energy models considered in this paper, using Eq. \eqref{eq:E2} we obtain:

\begin{equation}
\mathcal{H}_{0}(z) =H_0\frac{\sqrt{ \Omega_{m0} (1+z)^{3} + \Omega_{de}(z) }}{\sqrt{ \Omega_{m0} (1+z)^{3} + 1 - \Omega_{m0} }} \,.
\label{eq:Hrunning}
\end{equation}

\noindent In particular in \citet{Fazzari:2025mww, Montani:2025rcy} it has been shown that the $\mathcal{H}_0(z)$ diagnostic is able to discriminate between quintessence and phantom dark energy from respectively the increasing or decreasing behaviour of this function at low redshift.

\section{\label{sec:analysis}Data Analysis}

\subsection{\label{sec:dataset}Dataset: the Master binned sample}

\noindent The model will be fitted using a binned sample of SNe Ia compiled by \cite{Dainotti:2025qxz} and named \textit{Master binned sample}. Here, we briefly review the procedure implemented in \cite{Dainotti:2025qxz} to construct the data sample.

This dataset combines 3714 SNe Ia drawn from the Dark Energy Survey (DES)\citep{DES:2024jxu}, Pantheon+ \citep{Scolnic:2021amr, Brout:2022vxf}, Pantheon \citep{scolnic2018} and the Joint Lightcurve Analysis (JLA) \citep{JLA}, after eliminating duplicates. The contributions from each survey are summarized in Tab. \ref{tab:sn_sample}. We note that this sample has been constructed with a new statistical analysis that takes into account the most appropriate likelihoods following \cite{DAINOTTI202430}, which are not necessarily the Gaussian ones (see also \cite{Dainotti_2023, Bargiacchi_2023}).

\begin{table}[h!]
\centering
\caption*{\textbf{Master binned Sample}}
\vspace{-2mm} 
\begin{tabular}{l c}
\hline
Survey & Number of SNe Ia\\
\hline
\addlinespace[5pt]
DES & 1829 \\
Pantheon+ & 1208 \\
Pantheon & 181 \\
JLA & 496 \\
\addlinespace[5pt]
\hline
\end{tabular}
\vspace{1mm}
\caption{Sample of Type Ia Supernovae collected from various surveys to construct the Master Sample, after removing duplicate objects \citep{Dainotti:2025qxz}.}
\label{tab:sn_sample}
\end{table}

The resulting dataset is divided into $20$ bins of equal population, covering the range from $z=0.00122$ to $z=2.3$. The choice of $20$ bins aims for a balance between limiting the statistical uncertainties for each data point, by providing enough supernovae inside each bin, and keeping a sufficient number of data points for accurate fitting.
Within each interval, the observed distance modulus $\mu_{obs}$ is compared with the theoretical expectation that follows from a standard $\Lambda$CDM framework, whose luminosity distance is denoted as $d_L(z)$ : 

\begin{equation}
\mu_{\mathrm{obs}} = m_B^\ast - M_B + \alpha\,x_1 - \beta\,c + \Delta_M + \Delta_B,
\label{eq:mu_obs}
\end{equation}

\begin{equation}
\mu_{\mathrm{th}}(z) \;=\; 5\log_{10}\!\left(\frac{d_L(z)}{\mathrm{Mpc}}\right) + 25 \,,
\label{eq:mu_th}
\end{equation}

\noindent where $m_B^\ast$ is the observed rest-frame peak magnitude in the $B$ band, while $\alpha$ and $\beta$ are nuisance parameters that quantify the correlations between luminosity and, respectively, the light-curve shape parameter ($x_1$) and the color parameter ($c$). $\Delta_M$ is a distance correction accounting for the empirical tendency of SNe Ia in more massive galaxies to appear brighter, $\Delta_B$ corrects bias derived from survey simulations, and $M_B$ is the absolute B-band magnitude of a fiducial SN Ia
with $x_1 = 0$ and $c = 0$ (for details, see \cite{DAINOTTI202430}).

For a flat $\Lambda$CDM model, the luminosity distance is 

\begin{equation}
d_L(z) = (1+z)\frac{c}{H_0}\int_0^z\frac{\text{d}z'}{\sqrt{\Omega_{m0}(1+z')^3+(1-\Omega_{m0})}} \,,
\label{eq:dL_theo}
\end{equation}
depending inversely on $H_0$. Hence, to remove the degeneracy between $H_0$ and $M_B$ appearing in Eqs.~\eqref{eq:mu_th},~\eqref{eq:mu_obs},~\eqref{eq:dL_theo},
$M_B$ is calibrated using a reference value of $H_0=70$ km/s/Mpc, since it does not influence the qualitative descending trend of $\mathcal{H}_0(z)$. 

After the calibration procedure, the next step consists in obtaining the values of $H_0$ and $\Omega_{m0}$ in each redshift bin. These binned results are obtained through a Markov Chain Monte Carlo (MCMC) performed separately for each bin. Each redshift interval therefore provides a pair of estimates for $H_0$ and $\Omega_{m0}$, which are assumed to be statistically independent of those obtained in neighbouring bins.

Regarding the adopted priors, $H_0$ is assigned a uniform prior over the interval
$[60,80]~\mathrm{km\,s^{-1}\,Mpc^{-1}}$,
while the prior for $\Omega_{m0}$ is Gaussian, around the central value $\mathcal{N}\bigl(\mu=0.322,\sigma=0.025\bigr)$.
This value is obtained from a fit of the full sample within the $\Lambda$CDM model, with the prior width corresponding to the associated $5\sigma$ interval \citep{Dainotti:2025qxz}.

In this referenced paper, to assess the impact of neglecting possible inter-bin correlations, it was verified that the inferred best-fit parameters and their uncertainties remain stable when varying the binning scheme (e.g. changing the number of bins, implementing the moving window method, or using the $\log(z)$ division which takes into account the volumetric segregation). No statistically significant change was found in the results, indicating that residual inter-bin correlations do not materially affect the inferred uncertainties.

\subsection{\label{sec:fit}Fit procedure}

In this work, we perform an MCMC analysis using the \texttt{Cobaya} sampler \citep{cobaya}, interfaced with a custom Python code that numerically solves the system of equations in Eq.~\eqref{eq:E2} for each model. This allows us to reconstruct the corresponding theoretical running Hubble constant, defined in Eq.~\eqref{eq:Hrunning}, which is then compared with the one obtained from the Master Sample. We used four parallel chains with the Gelman-Rubin criterion $R-1<0.01$ \citep{gelman_rubin} to establish the convergence of the chains.

The parameter priors adopted are reported in Tab. \ref{tab:prior}. In particular, the priors on $H_0$ and $\Omega_{m0}$ are chosen to ensure consistency with the analysis of \citet{Fazzari:2025mww}, which includes the $H_\xi=0$ limit of our model (summarized in Section \ref{sec:nobulk}), thus allowing a comparison among all models. The prior on $H_\Gamma$ reflects our expectations for the matter creation rate in Sec. \ref{sec:constraints}, with a log-normal centered on $H_\Gamma \simeq 5$ and negligible weight below 1. 
Finally, statistical results and plots have been produced using \texttt{GetDist} \citep{getdist}. 

For model selection, we compute the differences in the Bayesian Information
Criterion (BIC) \citep{wagenmakers2007} for the three analysed models relative to the reference models, which are the $\Lambda$CDM and the Power-Law (PL) model. To interpret the significance of this factor, we adopt Jeffreys' scale \citep{Bayes_trotta, jeffreys_scale}, which categorizes the evidence against the model as inconclusive if $0 < \Delta \rm BIC < 1$, weak if $1 < \Delta \rm BIC < 2.5$, moderate if $2.5 < \Delta \rm BIC < 5.0$, and
strong if $\Delta \rm BIC > 5.0$.

\begin{table}[ht!]
\centering
\begin{tabular}{l l}
\hline
\addlinespace[2pt]
\textbf{Parameter} & \textbf{Prior} \\
\hline
\addlinespace[5pt]
$\Omega_{m0}$ & $U[0.01, 0.99]$ \\
$H_0$ (km/s/Mpc) & $U[60, 80]$ \\
$w$ & $U[-7, 3]$ \\
$H_\Gamma$ & $\operatorname{LogNormal}[1.6,0.5^2]$ 
\\
\addlinespace[5pt]
\hline
\end{tabular}
\caption{Prior distributions for the model parameters used in the MCMC procedure. $U[a,b]$ denotes a uniform prior between $a$ and $b$, while the prior on $H_\Gamma$ is log-normal with mean $\ln{H_\Gamma}=1.6$ and standard deviation $\sigma=0.5$. $H_\Gamma$ is sampled only for the complete model $M[\xi,\Gamma]$; $H_\xi$ is not sampled but derived from $q_0$, $\Omega_{m0}$, $w$ and $H_\Gamma$ through Eq.~\eqref{eq:H_xi_cosm}.}

\label{tab:prior}
\end{table}

We should again point out that, with this procedure, we are testing a new cosmological model using a dataset that was built assuming $\Lambda$CDM as the underlying model. This should be regarded as the first step of a possible more rigorous analysis, that would involve measuring the best-fit values of $\Omega_{m0}$, $H_0$ in each bin of Master Sample using the new model, instead of using the $\Lambda$CDM model. This first step is, however, useful: the real meaning of the claim that $\mathcal{H}_0(z)$ is a diagnostic tool lies in its strong capability of comparing the dynamical predictions of a proposed model with the standard $\Lambda$CDM case in a direct way (i.e. fitting the same collection of binned data points). By this tool, it is possible to discriminate between different cosmological models and select which ones seem to be significantly preferred with respect to the $\Lambda$CDM, and therefore which ones should be studied in more depth after possibly rebuilding the Master Sample with their revised dynamics. See \ref{app:A} for more details. 

\subsubsection{\label{sec:HgHxi} Preliminary considerations on $H_\Gamma$ and $H_\xi$}

We discuss here the prior assumptions adopted in the analysis for two phenomenological quantities $H_\Gamma$ and $H_\xi$.
First, the physical interpretation of the parameters $H_\Gamma$ and $H_\xi$ motivates us to restrict the analysis to positive values only. This additional condition is therefore imposed throughout the parameter exploration in the likelihood function.

Second, we should monitor the behaviour of the term $-H_\xi E$, which appears in the modified continuity equation Eq.~\eqref{eq:omegap} as the contribution associated with bulk viscosity. 
In a fluid description of the Universe, bulk viscosity accounts for deviations from thermodynamic equilibrium induced by the change in the volume of the system \citep{landau1981physical}. As any non-equilibrium effect, within a hydrodynamical description, the bulk viscosity contribution should therefore remain a perturbative correction with respect to the perfect-fluid contribution to the energy-momentum tensor. This requirement is essential in order to preserve a consistent physical interpretation of the model in the present framework, and can be expressed as $H_{\xi}E< |w|$. 

\begin{figure}[htpb]
\centering
  \includegraphics[width=1.0\linewidth]{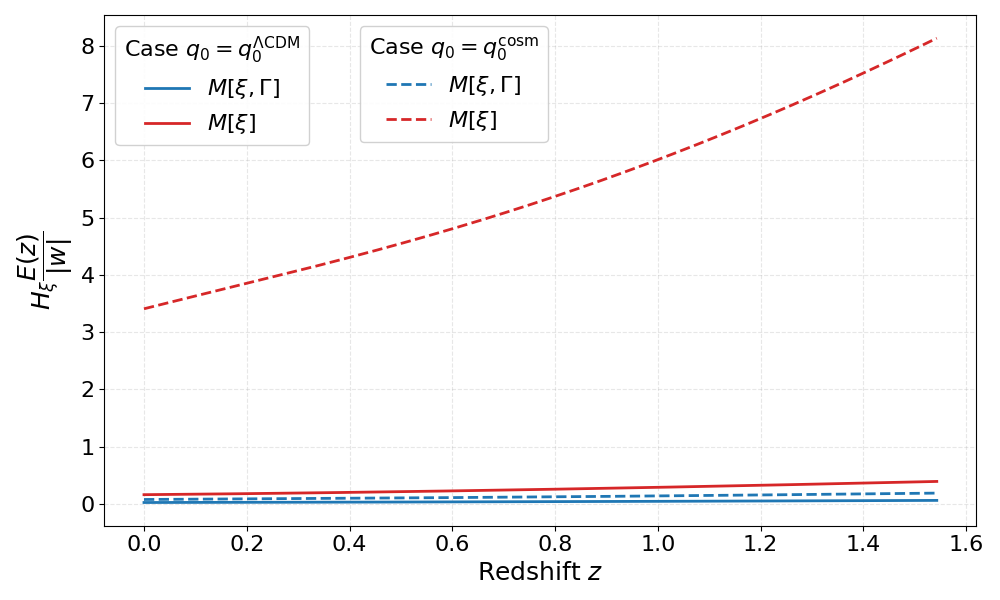}
  \caption{Check on the behaviour of the ratio $ \frac{|H_\xi E|}{|w|}$, evaluated using the fit parameters of Tab.~\ref{tab:results} for the two dynamical DE models involving bulk viscosity. In the studied redshift range, corresponding to the redshift range covered by the Master Sample data, if this term stays smaller than unity the perturbative character of the bulk viscosity is not violated. The dashed and continuous lines correspond to the two different assumptions on the deceleration parameter of each model discussed in Sec. \ref{sec:constraints}.}
\label{fig:bulkcheck}
\end{figure}

In Fig.~\ref{fig:bulkcheck}, we show the redshift evolution of this quantity over the redshift range covered by the data. The condition is satisfied for most of the cases considered, but it is violated in the model including only bulk viscosity when the deceleration parameter is fixed to $q_0=q_0^{\mathrm{cosm}}$. In this specific case, the model no longer fully realizes the physical scenario for which it was originally introduced. It should therefore be interpreted with caution, namely as an effective phenomenological parametrization rather than as a fully consistent bulk-viscous fluid model.

\subsubsection{Results and comparison}\label{sec:resultsandcomp} 
\begin{figure*}[ht!]
\centering
  \includegraphics[width=12cm]{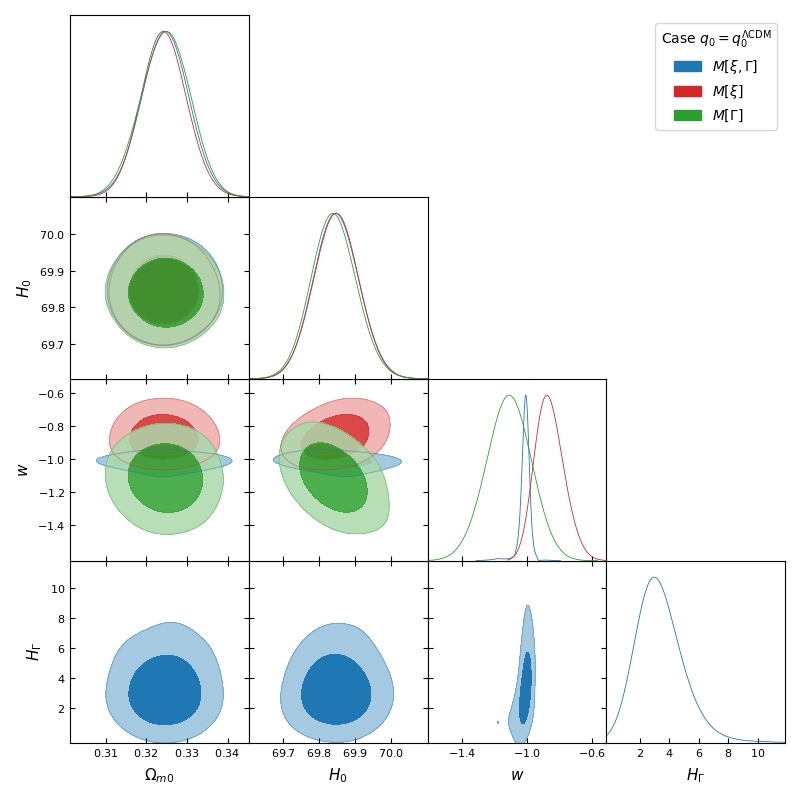}\\
  \includegraphics[width=12cm]{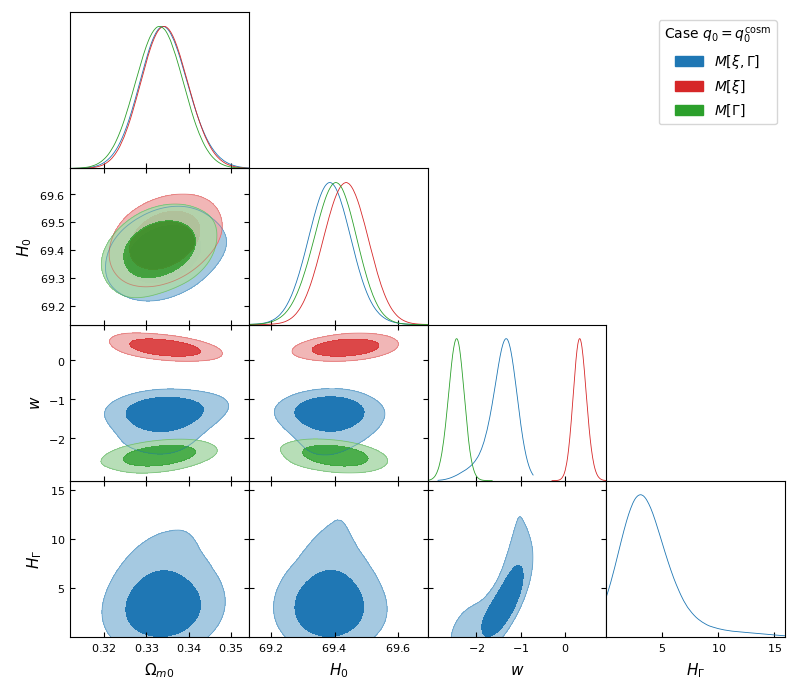}
\caption{Posterior probability distributions from the MCMC analysis of the Master binned Sample, for the free parameters of the three dynamical dark energy models $M[\xi,\Gamma]$, $M[\Gamma]$ and $M[\xi]$ ($68\%$ and $95\%$ confidence-level contours). The top and bottom panels correspond to the two different assumptions on the deceleration parameter of each model discussed in Sec. \ref{sec:constraints}.}
\label{fig:post}
\end{figure*}

\begin{figure*}[ht!]
\centering
  \includegraphics[width=12cm]{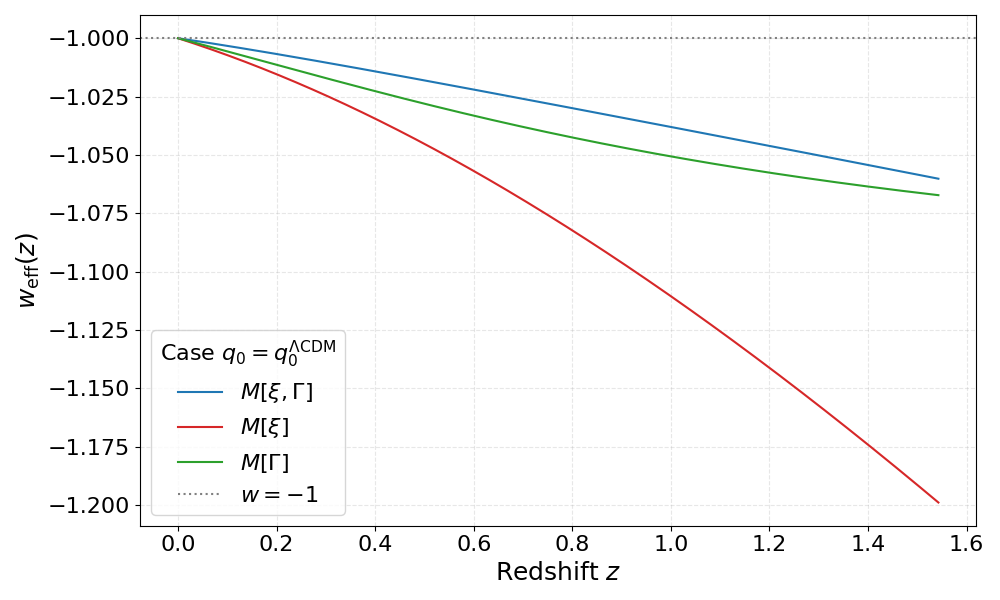}\\
  \includegraphics[width=12cm]{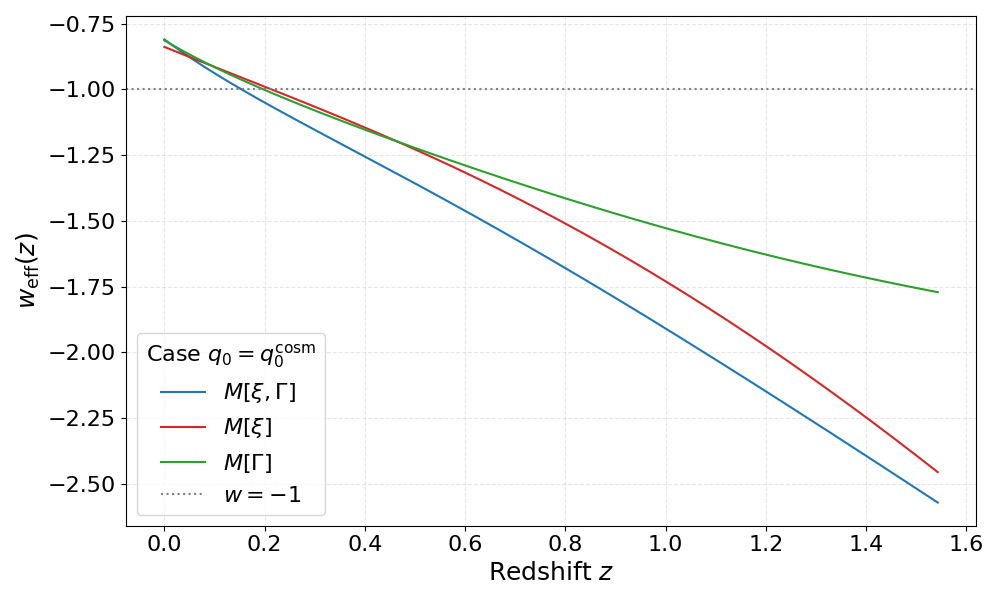}
\caption{Effective equation of state parameter for the three dynamical dark energy models, computed from the fitted parameters of Tab.~\ref{tab:results} using Eq.~\eqref{eq:weff} for $M[\xi,\Gamma]$, Eq.~\eqref{eq:weffmatter} for $M[\Gamma]$, and Eq.~\eqref{eq:weffbulk} for $M[\xi]$. The top and bottom panels correspond to the two different assumptions on the deceleration parameter of each model discussed in Sec. \ref{sec:constraints}.}
\label{fig:weff}
\end{figure*}

\begin{figure*}[ht!]
\centering
  \includegraphics[width=12cm]{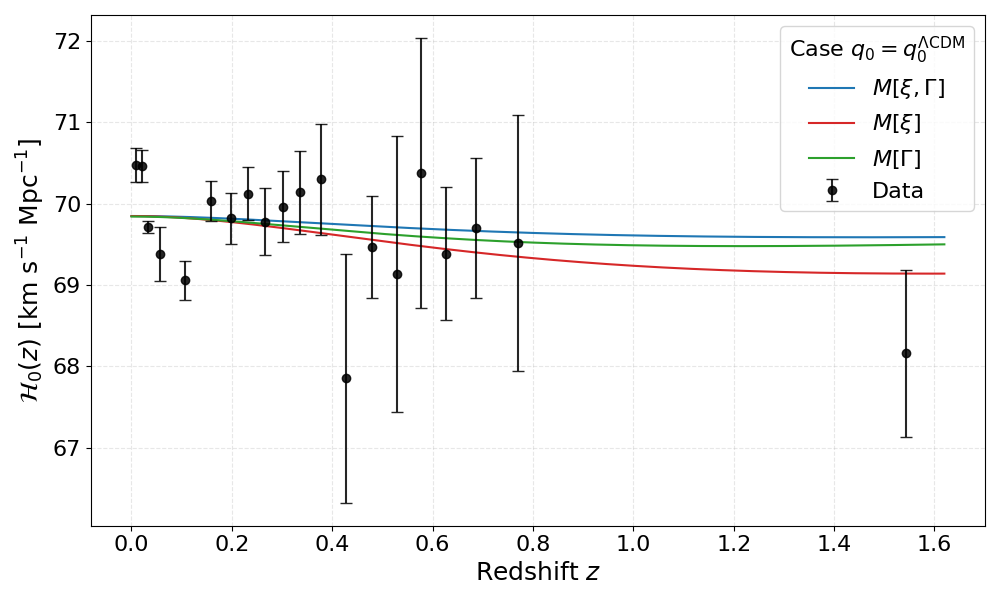}\\
  \includegraphics[width=12cm]{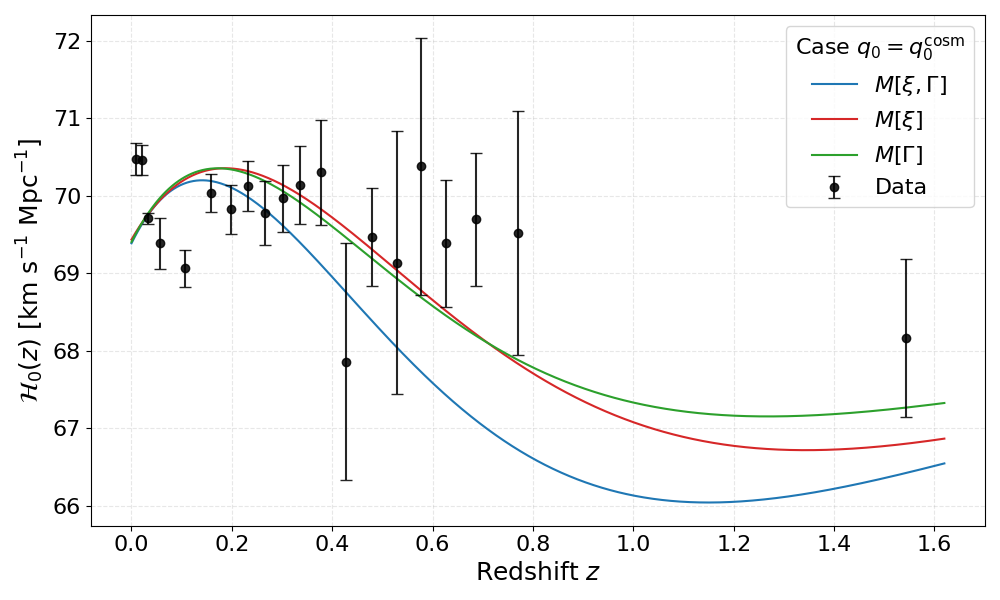}
\caption{Comparison between the theoretical effective running Hubble constant $\mathcal{H}_0(z)$, Eq.~\eqref{eq:Hrunning}, predicted by the three dynamical dark energy models $M[\xi,\Gamma]$, $M[\Gamma]$ and $M[\xi]$ (fitted parameters in Tab.~\ref{tab:results}), and the binned Master Sample data (Sec.~\ref{sec:dataset}). The top and bottom panels correspond to the two different assumptions on the deceleration parameter of each model discussed in Sec. \ref{sec:constraints}.}
\label{fig:fit}
\end{figure*}

We present here the results of the fitting procedure discussed in Sec.\ref{sec:fit}. The constraints on the model parameters are reported in Tab.~\ref{tab:results} for the different cosmological scenarios and for the two assumptions adopted for the present-day deceleration parameter, namely the $\Lambda$CDM low-redshift condition, $q_0=q_0^{\Lambda\mathrm{CDM}}$, and the cosmographic value, $q_0=q_0^{\mathrm{cosm}} $ (see Tab. \ref{tab:back_results}).
Fig. \ref{fig:post} shows the posterior distributions of the fitted parameters. The corresponding reconstructions of the effective equation of state parameter (EoS), $w_{\mathrm{eff}}(z)$, and of the effective running Hubble constant, $\mathcal{H}_0(z)$, are shown in Figs \ref{fig:weff}-\ref{fig:fit}. Finally, we show in Tab. \ref{tab:bic-combined} the resulting BIC values of each model; the BIC relating to the $\Lambda \mathrm{CDM}$ fit and Power-Law (PL) fit of the same dataset is also displayed for reference. We remind here that the PL model is a phenomenological function of the form $\mathcal{H}_0(z)=H_0(1+z)^{-\alpha}$, $\alpha=0.01 \pm 0.008$, $H_0=69.87$ km/s/Mpc (see Table 6 of  the Master sample analysis \citep{Dainotti:2025qxz}), which has so far been statistically favoured by the SNe Ia data when compared to various dynamical DE models (see \citet{Fazzari:2025mww}), and is the only one comparable to $\Lambda \mathrm{CDM}$, which still holds the lowest BIC value.

\begin{table*}[htbp]
\centering
\renewcommand{\arraystretch}{1.35}
\begin{tabular}{cccc}
\hline\hline
\multicolumn{4}{c}{\textbf{Master binned}}\\
\hline\hline
\textbf{$q_0^{\mathrm{cosm}}$} & $M[\xi,\Gamma]$ & $M[\xi]$ & $M[\Gamma]$ \\[2pt]
\hline
$H_0$ [km\,s$^{-1}$\,Mpc$^{-1}$] & $69.38 \pm 0.06$ & $69.43 \pm 0.06$ & $69.40 \pm 0.06$ \\
$\Omega_{m0}$             & $0.3340 \pm 0.0053$ & $0.3343 \pm 0.0053$ & $0.3330 \pm 0.0053$ \\
$w$                    & $-1.42 \pm 0.30$ & $0.32 \pm 0.15$ & $-2.45 \pm 0.17$ \\
$H_\xi$                & $0.46 \pm 0.14$ & $1.16 \pm 0.15$ & -- \\
$H_\Gamma$             & $3.1 \pm 1.5$ & -- & $1.130 \pm 0.015$ \\
\hline
$\chi^2_{\mathrm{red}}$ & $2.83$ & $2.26$ & $2.28$ \\
\hline\hline
\textbf{$q_0^{\Lambda \mathrm{CDM}}$} & $M[\xi,\Gamma]$ & $M[\xi]$ & $M[\Gamma]$ \\[2pt]
\hline
$H_0$ [km\,s$^{-1}$\,Mpc$^{-1}$] & $69.85 \pm 0.06$ & $69.85 \pm 0.06$ & $69.84 \pm 0.06$ \\
$\Omega_{m0}$             & $0.3246 \pm 0.0055$ & $0.3243 \pm 0.0052$ & $0.3246 \pm 0.0055$ \\
$w$                    & $-1.013 \pm 0.029$ & $-0.86 \pm 0.085$ & $-1.11 \pm 0.13$ \\
$H_\xi$                & $0.015 \pm 0.010$ & $0.145 \pm 0.087$ & -- \\
$H_\Gamma$             & $2.7 \pm 1.2$ & -- & $1$ \\
\hline
$\chi^2_{\mathrm{red}}$  & $1.18$ & $1.14$ & $1.15$ \\
\hline\hline
\end{tabular}
\caption{Mean fitted parameters resulting from the MCMC analysis for the three dark energy models, using the Master binned Sample and the priors in Tab. \ref{tab:prior}. $H_0, \, w$ and $\Omega_{m0}$ are free parameters; $H_\Gamma$ is a free parameter only for the complete model M[$\xi,\Gamma$], while it equals $1$ for the $M[\Gamma]$ model in the $\Lambda\mathrm{CDM}$ limit as indicated in Eq. \eqref{eq:H_g_LCDM_1}; $H_\xi$ is a derived parameter for all models (see Eq. \eqref{eq:H_xi_cosm}). The symbol ``--'' indicates that the corresponding parameter is set to zero by construction in that specific model. The top and bottom blocks correspond to the two different assumptions on the deceleration parameter of each model discussed in Sec. \ref{sec:constraints}.}
\label{tab:results}
\end{table*}

Looking at Tab. \ref{tab:results}, we can firstly observe that the results emerging from the fit strongly depend on the chosen $q_0$. In the case of a cosmographic $q_0$, i.e. $q_0=q_0^{\rm cosm}$, the bulk viscosity only model $M[\xi]$ results in a positive, radiation-like value for $w$ (though this case violates the perturbative condition), while the matter creation only model $M[\Gamma]$ selects a phantom character; when we consider the combined effects in the complete model  $M[\xi, \Gamma]$, they seem to balance into a milder (closer to $-1$) phantom value of $w$.

An analogous outcome can be seen in the $\Lambda \mathrm{CDM}$ limit. Since $w_{\mathrm{eff}}$ needs to enter the phantom regime in order for $\mathcal{H}_0(z)$ to show the slight descending trend in the data (see Fig. \ref{fig:weff}), the constraint on $q_0$ of Eq. \eqref{eq:q0_LCDM} forces $w>-1$ for the bulk viscosity only model $M[\xi]$ and $w<-1$ for the matter creation one $M[\Gamma]$ (otherwise their functions $w_{\mathrm{eff}}$, following Eqs. \eqref{eq:q0lcdmweffbulk} \eqref{eq:q0lcdmweffmatter} respectively, could not drop below $-1$). More details on the relation between $w_{\mathrm{eff}}$ and the decreasing or increasing trend for $H_0(z)$ can be found in  \citet{Fazzari:2025mww}.

Secondly, the observational data constrain the fitted values of $\Omega_{m0}$ close to
$0.33$, so that $q_0^{\Lambda\mathrm{CDM}}\simeq-0.50$; since the cosmographic values
obtained in \ref{app:background_preliminary} (see Tab.~\ref{tab:back_results}) satisfy
$q_0^{\mathrm{cosm}}>q_0^{\Lambda\mathrm{CDM}}$ for all three models, the function
$A(q_0,\Omega_{m0})$ is positive (see Eq.~\eqref{eq:H_xi_cosm}).

The poor constraining power on $H_\Gamma$ in $M[\xi,\Gamma]$ under both cases has a structural origin that is worth examining in detail.

Judging by the posteriors in Fig. \ref{fig:post} and the strong degeneracy between the fitted parameters $w$ and $H_\Gamma$, what the data actually constrain is the term $(w+1)(1-H_\Gamma)$ since it is directly linked to $H_\xi$ (Eqs. \eqref{eq:H_xi_cosm}, \eqref{eq:q0lcdmconstrain}), and therefore to the drift term $-H_\xi E$ appearing in the effective function $w_{\mathrm{eff}}$ (Eq. \eqref{eq:weff}). This is the only term giving a negative contribution to $w_{\mathrm{eff}}$, while the sign of the matter creation term $(1+w)(1-\frac{H_\Gamma}{E(z)})$ stays positive in the whole range of the data; therefore, $H_\Gamma$ does not contribute to the quality of fit, instead worsening it by counteracting the bulk viscosity effect and opposing the desired phantom trend.
This is also why, in the $q_0=q_0^{\Lambda\mathrm{CDM}}$ case, $w$ approaches $-1$ for large values of $H_\Gamma$  so as to neutralize the matter-creation contribution  $(w+1)(1-H_\Gamma)$. This behaviour is clearly visible in Fig.~\ref{fig:post}, where the constraint on $w$ for the complete model $M[\xi,\Gamma]$ is tighter and centered around $-1$. The reason why this does not happen in the $q_0=q_0^{\mathrm{cosm}}$ case is that $w$ must be pushed sufficiently far from $-1$ in order to satisfy the requirement $H_\xi>0$ with a positive $A(q_0,\Omega_{m0})$ (see Eq. \eqref{eq:H_xi_cosm}).

This should not be interpreted as evidence that matter creation is generally disfavoured when the goal is to alleviate the Hubble tension in the late Universe. Rather, this conclusion holds only under the specific assumptions adopted in the present analysis, which strongly restrict the role of matter creation, namely the requirement $H_\xi>0$, the fixing of $q_0$ to the adopted values, and the assumption of a constant creation rate $\Gamma_0$ introduced in Sec. \ref{sec:model}.

Looking at the effective EoS parameter in Fig. \ref{fig:weff}, 
the models with $q_0=q_0^{\mathrm{cosm}}$ show a quintessence-to-phantom transition around $z_c \sim 0.2$. In fact, $w_{\mathrm{eff}}(z=0)=-1+A(q_0,\Omega_{m0})>-1$ since $A$ is positive for all models, as explained above.

As expected from the previous analysis of \citet{Fazzari:2025mww}, the reconstructed function $\mathcal{H}_0(z)$ can be used as a diagnostic of the DE regime: at low redshift, a positive slope is associated with quintessence-like behaviour, while a negative slope indicates a phantom-like behaviour (it can be observed in Fig.~\ref{fig:fit} that $z_c$ coincides at leading order with the maximum point for  $\mathcal{H}_0(z)$).\footnote{This can be understood as follows. $\mathcal{H}_0^2(z)\propto 1+\delta \Omega(z)/E^2_{\Lambda\rm CDM}(z)$, with $\delta \Omega(z) \equiv \Omega_{de}(z)-\Omega_{\Lambda}$ measuring the departure of the dark energy density from its
$\Lambda$CDM value, and $E^2_{\Lambda\rm CDM}(z)\equiv\Omega_{m0}(1+z)^3+\Omega_{\Lambda}$. The sign of $\mathcal{H}_0'(z)\equiv\mathrm{d}\mathcal{H}_0(z)/\mathrm{d}z$ is then set by a competition between two terms, a push proportional to the derivative of $\delta{\Omega}$ (see Eq. \eqref{eq:omegap}), and a drag proportional to $\delta\Omega$ itself:
\begin{equation}
\mathcal{H}_0'(z)\propto \underbrace{E^2_{\Lambda\rm CDM}\,(1+w_{\mathrm{eff}})\,\Omega_{de}}_{\text{push}}-\underbrace{\delta\Omega\,\,\Omega_{m0}(1+z)^3}_{\text{drag}}\,.
\end{equation}
By construction $\delta\Omega(0)=0$, so at low redshift the drag term is still negligible and the sign of $\mathcal{H}_0'(z)$ is governed by the push alone. Since $\delta\Omega$ remains small up to $z_c$, this holds in particular at the transition, where $w_{\mathrm{eff}}(z_c)=-1$ makes the push vanish and therefore $\mathcal{H}_0(z)$ reaches the maximum.}

We should take into account, however, that the quality of fit and BIC value (in Tab. \ref{tab:bic-combined}) improves significantly for all models when $q_0=q_0^{\Lambda\mathrm{CDM}}$, i.e. when no transition is possible.

Moreover, in the case of $q_0=q_0^{\rm cosm}$ another problem occurs for the bulk viscosity only scenario $M[\xi]$. Starting from a quintessence value of $w_{\mathrm{eff}}$ at $z=0$ (as a consequence of having considered the cosmographic $q_0$) forces the bulk viscosity to become non-perturbative (assuming a large value of $H_\xi$) since a stronger negative pressure is needed in order to lower $w_{\mathrm{eff}}(z)$ from the initial starting point above $-1$ to the effective phantom regime, as the redshift increases.

\begin{table}[htbp]
\centering
\renewcommand{\arraystretch}{1.35}
\begin{tabular}{lccc}
\hline\hline
\multicolumn{4}{c}{\textbf{BIC values}} \\
\hline

& $M[\xi,\Gamma]$ 
& $M[\xi]$ 
& $M[\Gamma]$ \\
\hline

$q_0^{\mathrm{cosm}}$ 
& $101.23$ & $94.77$ & $95.62$ \\

$q_0^{\Lambda\mathrm{CDM}}$ &
$57.10$ & $53.23$ & $53.66$ \\

\hline
\multicolumn{4}{c}{Reference models} \\
\hline

$\Lambda\mathrm{CDM}$ 
& \multicolumn{3}{c}{$46.24$} \\

PL 
& \multicolumn{3}{c}{$46.91$} \\

\hline\hline
\end{tabular}
\caption{Bayesian Information Criterion (BIC) for the three dynamical dark energy models fitted to the Master binned Sample, under the two assumptions for the present-day deceleration parameter $q_0$. The lower block reports the BIC values for the reference $\Lambda$CDM and power-law (PL) models ($\mathcal{H}_0(z) = H_0(1+z)^{-\alpha}$ with $\alpha = 0.01$ and $H_0 = 69.869\ \mathrm{km\,s^{-1}\,Mpc^{-1}}$). The BIC is evaluated at the maximum-likelihood sample of each chain. A lower BIC value corresponds to a better quality of fit after penalizing each model for its complexity (i.e. its number of degrees of freedom): for all three dynamical models, the fit obtained under $q_0 = q_0^{\Lambda\mathrm{CDM}}$ is strongly preferred over the one obtained under $q_0 = q_0^{\mathrm{cosm}}$ according to Jeffreys' scale \citep{Bayes_trotta, jeffreys_scale}.}

\label{tab:bic-combined}
\end{table}

\section{Conclusions}\label{sec:conclusion}
In this work, we investigated an extension of the standard $\Lambda$CDM scenario in which the dark energy (DE) component is subject to both particle creation by the expanding gravitational field and non-equilibrium effects described through a bulk-viscosity contribution. The resulting framework naturally leads to a dynamical DE sector characterized by an evolving effective equation of state and by a modified expansion history; following previous studies, we employed the effective running Hubble constant as a diagnostic tool to compare the predictions of the proposed models with the behaviour inferred from the binned Type Ia Supernovae (SNe Ia) Master Sample.

Since the deceleration parameter $q_0$ enters directly into the constraint relating the model's parameters $H_{\xi}$, $H_{\Gamma}$ and $w$, we assessed the robustness of the inferred DE properties under different assumptions on $q_0$. To this end, we repeated the analysis for all three dynamical dark-energy models considered in this work, namely the complete model $M[\xi, \Gamma]$, the matter-creation model $M[\Gamma]$, and the bulk-viscosity model $M[\xi]$, using both $q_0=q_0^{\Lambda{\mathrm{CDM}}}$ and the cosmographic $q_0=q_0^{\mathrm{cosm}}$ as constrained in \ref{app:background_preliminary}. 

Firstly, we found that the complete model and the matter creation only model favour an intrinsic EoS parameter in the phantom regime, while the bulk viscosity only model is associated with a quintessence-like component. Under both choices of $q_0$, matter creation does not provide a useful dynamical contribution to the fit: as shown in detail in Sec.~\ref{sec:resultsandcomp}, the drift term $-H_\xi E$ is the only channel that can drive the required phantom trend in $w_{\mathrm{eff}}(z)$, while the creation term stays positive throughout the data range and effectively opposes it. As a result, the complete model $M[\xi,\Gamma]$ does not improve over its one-parameter limits and is penalized by the BIC.

Secondly, the comparison sheds light on the validity of the perturbative bulk-viscosity scenario. For the bulk viscosity only model $M[\xi]$, adopting the cosmographic value $q_0=q_0^{\mathrm{cosm}}$ leads to values of $H_\xi E(z)$ that exceed the perturbative regime over the entire investigated redshift range (this case also finds a radiation-like value for $w$, not compatible with dark energy). On the other hand, when $q_0=q_0^{\Lambda\mathrm{CDM}}$ is imposed, the perturbative condition remains roughly satisfied ($H_\xi E(z)/|w|<0.4$), showing that the bulk viscosity only scenario cannot be excluded \textit{a priori} and that the apparent inconsistency is conditional on the choice of deceleration parameter rather than intrinsic to the viscosity framework itself.

Another result of the present analysis is that the inferred behaviour of DE depends on the adopted value of the present-day deceleration parameter. When the cosmographic value of $q_0$ is adopted, the effective EoS of the three models exhibits a transition from a quintessence-like phase to a phantom phase at redshift $z_c \sim 0.2$. The transition is instead not permitted when the $\Lambda$CDM value of $q_0$ is imposed, in which case the quality of fit also improves for all models ($\Delta\mathrm{BIC} > 40$).  
Our conclusion, however, is restricted to the chosen models of this study, and therefore it should not be taken as decisive evidence against a quintessence to phantom transition in the SNe Ia data. 

The main finding of the present work is not that a specific dynamical scenario is uniquely preferred, but rather that the inferred character of DE is highly sensitive to the assumed physical and cosmological constraints of the chosen model. This result highlights the importance of combining model-independent approaches with physically motivated dynamical models when investigating the nature of DE and the origin of the Hubble tension. Future analyses based on substantially larger Supernova samples, together with independent late-time probes such as BAO and cosmic chronometers, will be essential to establish whether the phantom transition suggested by several recent studies is a genuine physical feature of DE or a consequence of the assumptions adopted in its reconstruction.

\section*{CRediT authorship contribution statement}

\textbf{I. Navone:} Software, Methodology, Formal Analysis, Writing - Original Draft, Writing - Review \& Editing, Visualization. \textbf{M. G. Dainotti:} Data curation, Supervision, Writing - Original Draft, Writing - Review \& Editing, Conceptualization, Methodology, Investigation, Formal analysis. \textbf{E. Fazzari:} Software supervision, Formal Analysis, Writing - Review \& Editing. \textbf{G. Montani:} Conceptualization, Methodology, Writing - Original Draft, Writing - Review \& Editing. \textbf{N. Maki:} Writing - Review \& Editing, Validation. \textbf{K. Kohri:} Conceptualization.   

\section*{Declaration of competing interest}

The authors declare that they have no known competing financial
interests or personal relationships that could have appeared to influence the work reported in this paper. 

\section*{Data availability}

Data will be made available upon reasonable request to the corresponding author.

\section*{Acknowledgments}
M.G. Dainotti acknowledges the support of the DoS for her travel to La Sapienza in February 2026 and in July 2026. M.G.D. acknowledges the support of the JSPS Grant-in-Aid for Scientific Research (KAKENHI) (A), Grant Number JP25H00675 for supporting her travel and accommodation to visit La Sapienza in September 2025 and in July 2026. I.N. acknowledges the support of DoS via the Exploratory Research Grant 2025, the support of the Asian Winter School funded by Sokendai, the support of NAOJ and the hospitality of Cosmo Lodge during the school. 
E.F. is supported by ``Theoretical Astroparticle Physics'' (TAsP), iniziativa specifica INFN. 

\newpage

\appendix
\section{The validity of our diagnostic tool within the simple power-law model}\label{app:A}
Ideally, each new model should be implemented to construct the function $\mathcal{H}_0(z)$, and the model which best fits the data should be identified as the one for which such a function best resembles, from a statistical point of view, a constant function of the redshift.
However, such a procedure requires strong constraining power from the binning representation, i.e. a very large number of SNe Ia in each bin. This will become possible in 5-10 years, when according to the upcoming observations the number of known SNe Ia will grow by a factor of 100 (including high redshift ones).

At the present stage of available data, the implementation of a model having many free parameters cannot always be constrained in a statistically reliable way. Thus, we introduced a diagnostic tool (i.e. a theoretical function) which allows us to estimate the improvement that models are able to produce with respect to the standard $\Lambda$CDM model: such a capability is measured by the statistical quality (provided by this diagnostic tool) of the fit to the data coming from a $\Lambda$CDM binned analysis. This methodology allows one to perform a wide preliminary screening among many possible models, providing a number of candidates to be directly tested in the future once the statistics is enlarged.
Indeed, the models are numerous and before focusing on a particular subset, it would be relevant to select the most promising ones.

Up to now, we considered dynamical DE models which, unlike the $\Lambda$CDM one, are able to 
produce a non-constant $\mathcal{H}_0(z)$ and can therefore accommodate the mild 
decreasing trend shown by the binned data, although none of them reproduces it as 
effectively as the phenomenological power-law decreasing effective running Hubble 
constant, introduced in \citet{dainotti2021, dainotti2022, Dainotti:2025qxz} and 
physically motivated in \citet{Montani_carlevaro_dainotti}.
We here show that our methodology works well by implementing this best performing model directly. Thus, we show here a study of the Master binned Sample data via the power-law model, which is \textit{de facto} a conformal scaling of the $\Lambda$CDM one via $(1+z)^{-\alpha}$.

For each bin, we applied the same procedure described in Sec. \ref{sec:dataset}, after modifying the luminosity distance function by assuming the power-law dynamics rather than the $\Lambda \mathrm{CDM}$ dynamics. This led to a new data sample that we fitted with the power-law function $\mathcal{H}_0(z)=H_0(1+z)^{-\alpha}$. The results of this new study show that the emerging running Hubble constant is now more compatible with a real constant: the resulting data yield a fitted $\alpha=0.024 \pm 0.025$, compatible with 0 within $0.97\,\sigma$, while the original binned sample (built with the $\Lambda$CDM luminosity distance) yielded a fitted $\alpha$ differing from 0 by 1.3 $\sigma$ (see Tab. 6 of \cite{Dainotti:2025qxz}). This analysis has been performed with a uniform prior on $H_0$ and a Gaussian prior on
$\Omega_M$ whose width corresponds to a $5\,\sigma$ interval, following the strategy adopted
for the Master Sample in \cite{Dainotti:2025qxz}. In this case, however, the prior on
$\Omega_M$ is centred at $\Omega_M=0.35$ with $5\sigma=0.085$: this central value is
obtained by running the analysis on the full Master Sample with the modified luminosity
distance, adopting uniform priors $H_0\in U[55,95]\,\mathrm{km\,s^{-1}\,Mpc^{-1}}$ and
$\Omega_M\in U[0.01,0.99]$, together with a Gaussian prior on $\alpha$ centred at $0.01$
with $5\sigma=0.04$. We then take the average of the best-fit values obtained from two such
runs, performed respectively with the best-fit likelihoods and with the Gaussian likelihoods
of the full Master Sample; for the difference between the two treatments, see \citet{Dainotti:2025qxz}, \cite{dainotti2024reduceduncertainties43hubble}.

Had the diagnostic failed, the reconstructed $\mathcal{H}_0(z)$ would not have shown a trend more compatible with a flat one than the $\Lambda$CDM-based sample found. This additional analysis illustrates the internal consistency of the diagnostic
framework and supports its use as a preliminary model-selection tool. We stress that this should be
read as an argument in favour of the methodology rather than as a definite proof of its general validity,
consistently with the exploratory purpose for which the diagnostic was introduced: to screen candidate models, rather than to select among them conclusively.

\section{Background preliminary analysis}\label{app:background_preliminary}
\begin{figure}[htpb]
    \centering
    \includegraphics[width=1.0\linewidth]{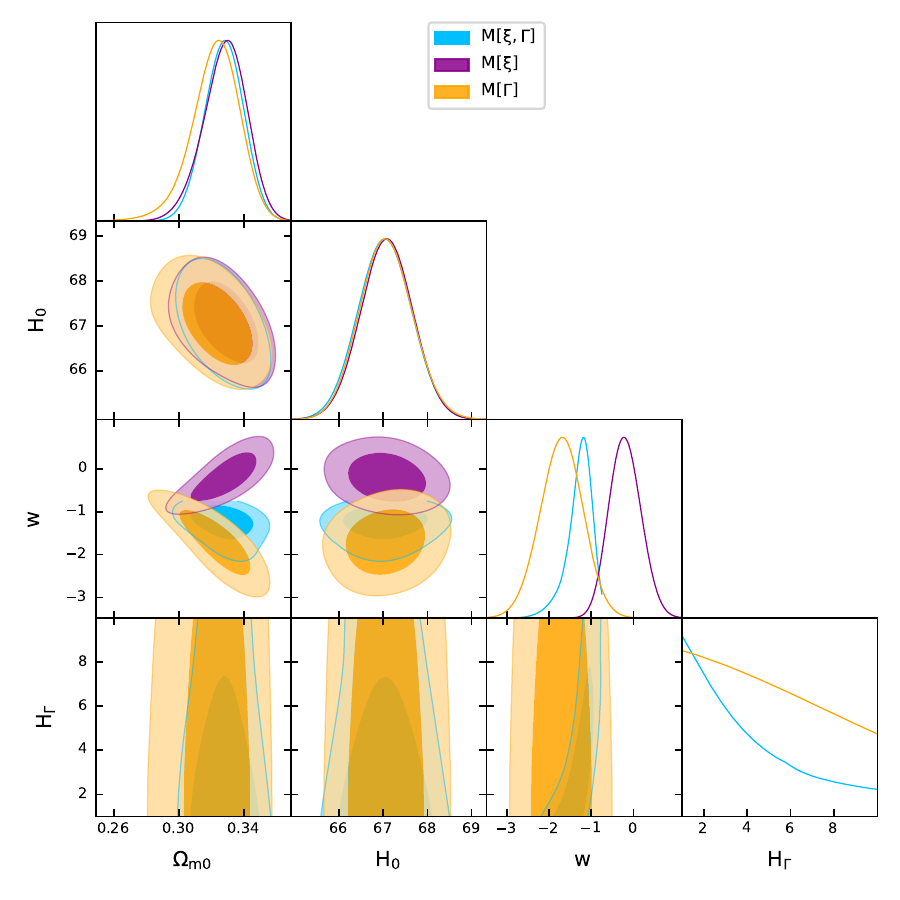}
    \caption{One-dimensional posterior probability distributions and two-dimensional $68\%$ and $95\%$ confidence level contours on cosmological parameters, obtained from the background analysis adopting the Gaussian cosmographic prior on $q_0$. The posterior distribution of $w$ in the complete model is truncated for $w\gtrsim -0.9$ because, after imposing the constraint on $q_0$, $H_\xi$ becomes a derived parameter through Eq.~\eqref{eq:H_xi_cosm}. In this region, the combination of the adopted prior range for $H_\Gamma$ and the physical requirement $H_\xi\geq0$ excludes part of the parameter space.}
    \label{fig:background_cosm}
\end{figure}

In this appendix, we present a preliminary background level analysis of the three dynamical dark energy models. The purpose of this preliminary step is to obtain reference confidence regions for the model parameters, which can then be used to guide the choice of prior ranges in the subsequent binned SNe Ia analysis, and more importantly to infer a cosmographic value of $q_0$ for each model.
We consider the joint combination of the following background probes:
\begin{itemize}
    \item SNe Ia Master Sample: a combined Type Ia supernova compilation obtained by merging DESY5, PantheonPlus, Pantheon and JLA, after removing duplicated objects \citep{Dainotti:2025qxz}. We refer to this dataset as \textit{Master}.
    \item DESI-DR2 BAO: baryon acoustic oscillation measurements from the second-year DESI data release \citep{DESI:2025zgx}, covering the redshift range $0.1<z<4.2$. The BAO observables are calibrated using the Planck value of the sound horizon,
    $r_d = (147.09 \pm 0.26)\,\mathrm{Mpc}$, as reported in Tab. 2 of Ref.~\citet{Planck2018}. We denote this dataset as \textit{DESI}.
    \item Cosmic Chronometers: direct measurements of the expansion rate $H(z)$ obtained from the differential-age method applied to passively evolving galaxies \citep{Jimenez_247, CC_borghi}. In this analysis, we use a subset of 15 measurements from Refs.~\citet{Moresco_2012, moresco_2015, Moresco_2016}, spanning the redshift interval $0.1791<z<1.965$. This dataset is denoted as \textit{CC}.
\end{itemize}
Moreover, we used the deceleration parameter from the result of the cosmographic, model-independent analysis performed in Ref.~\citet{Fazzari:2025lzd}. In particular, we adopt the value $q_0^{\mathrm{cosm}}=-0.366\pm 0.225$ from Tab. 1 of that work, extending the error to $5\sigma$, corresponding to the DESI+DESY5+CC data combination and to the same DESI-BAO calibration based on the Planck sound horizon. This choice is motivated by the fact that DESY5 represents the dominant SNe Ia contribution to the Master Sample, while the DESI and CC datasets coincide with those used in the present analysis. \\
As in the main analysis, we sample the posterior distribution using the MCMC algorithm implemented in \texttt{Cobaya} \citep{cobaya}, and we assess chain convergence through the Gelman-Rubin criterion $R-1 <0.01$ \citep{gelman_rubin}. We adopt the following uniform priors $H_0 \in \mathcal{U}[60,80]\,\mathrm{km\,s^{-1}\,Mpc^{-1}}, \; \Omega_{m0} \in \mathcal{U}[0.01,0.99], \; w \in \mathcal{U}[-7,3],\; H_{\Gamma} \in \mathcal{U}[1,10]$, motivated by the theoretical constraints discussed in Section~\ref{sec:constraints}. Constraints on $H_{\xi}$ and $q_0$ are obtained as derived parameters using the expressions derived in \autoref{sec:constraints}. \\
Fig. \ref{fig:background_cosm} shows the posterior contour plots of the cosmological parameters for the complete model $M[\xi,\Gamma]$, the bulk viscosity only model $M[\xi]$ and the matter creation only model $M[\Gamma]$ when using a Gaussian prior on $q_0=q_0^{\mathrm{cosm}}$. Tab.~\ref{tab:back_results} instead reports the mean values and the corresponding $68\%$ confidence-level intervals. For the complete and for the matter creation only models the fit provides only an upper limit for the $H_{\Gamma}$ value, showing that the maximum of the posterior probability is around $H_{\Gamma}=1$. We do not extend the prior range to lower values due to the theoretical constraints $H_{\Gamma}>1$ and $H_{\xi}>0$ discussed in Secs. \ref{sec:constraints}, \ref{sec:HgHxi} and imposed in the likelihoods.

\begin{table}[htbp]
\centering
\footnotesize
\setlength{\tabcolsep}{3pt}
\renewcommand{\arraystretch}{1.25}

\begin{tabular}{@{}lccc@{}}
\hline\hline
\multicolumn{4}{c}{\textbf{Master + DESI + CC}}\\
\hline\hline
\textbf{$q_0^{\mathrm{cosm}}$}
& $M[\xi,\Gamma]$
& $M[\xi]$
& $M[\Gamma]$\\
\hline
$H_0$ [km/s/Mpc]
& $67.04 \pm 0.57$
& $67.08 \pm 0.55$
& $67.06 \pm 0.56$ \\

$\Omega_{m0}$
& $0.328 \pm 0.011$
& $0.328^{+0.014}_{-0.012}$
& $0.323^{+0.015}_{-0.013}$ \\

$w$
& $-1.28^{+0.34}_{-0.19}$
& $-0.20 \pm 0.35$
& $-1.70 \pm 0.47$ \\

$H_\Gamma$
& $<4.82$
& --
& $<1.34$ \\

$H_\xi$
& $0.22^{+0.14}_{-0.17}$
& $0.63 \pm 0.31$
& -- \\

$q_0$
& $-0.309\pm0.067$
& $-0.337\pm 0.061$
& $-0.314\pm 0.077$ \\
\hline\hline
\end{tabular}

\caption{Mean values and corresponding $68\%$ confidence-level intervals obtained from the background analysis for the complete model $M[\xi,\Gamma]$, the bulk viscosity only model $M[\xi]$, and the matter creation only model $M[\Gamma]$, considering the cosmographic prior for the present-day deceleration parameter. Upper and lower limits are reported when the corresponding parameter is not constrained within the explored prior range. The corresponding values of $H_{\xi}$ and $q_0$ are derived according to the assumptions adopted for each model.}
\label{tab:back_results}
\end{table}

\clearpage

\bibliographystyle{elsarticle-harv} 
\bibliography{references}

@article{DESI:2024mwx,
    author = "Adame, A. G. and others",
    collaboration = "DESI",
    title = "{DESI 2024 VI: cosmological constraints from the measurements of baryon acoustic oscillations}",
    eprint = "2404.03002",
    archivePrefix = "arXiv",
    primaryClass = "astro-ph.CO",
    reportNumber = "FERMILAB-PUB-24-0154-PPD",
    doi = "10.1088/1475-7516/2025/02/021",
    journal = "JCAP",
    volume = "02",
    pages = "021",
    year = "2025"
}

@misc{dainotti2024reduceduncertainties43hubble,
      title={Reduced uncertainties up to 43{\%} on the Hubble constant and the matter density with the SNe Ia with a new statistical analysis}, 
      author={Maria Giovanna Dainotti and Giada Bargiacchi and Malgorzata Bogdan and Salvatore Capozziello and Shigehiro Nagataki},
      year={2024},
      eprint={2024JHEAp..41...30D},
      archivePrefix={arXiv},
      primaryClass={astro-ph.CO},
      url={https://arxiv.org/abs/2303.06974}, 
}

@article{Dainotti_2023,
   title={Reducing the Uncertainty on the Hubble Constant up to 35{\%} with an Improved Statistical Analysis: Different Best-fit Likelihoods for Type Ia Supernovae, Baryon Acoustic Oscillations, Quasars, and Gamma-Ray Bursts},
   volume={951},
   ISSN={1538-4357},
   url={http://dx.doi.org/10.3847/1538-4357/acd63f},
   DOI={10.3847/1538-4357/acd63f},
   number={1},
   journal={The Astrophysical Journal},
   publisher={American Astronomical Society},
   author={Dainotti, Maria Giovanna and Bargiacchi, Giada and Bogdan, Malgorzata and Lenart, Aleksander Lukasz and Iwasaki, Kazunari and Capozziello, Salvatore and Zhang, Bing and Fraija, Nissim},
   year={2023},
   month={jul}, pages={63} }

@misc{leclair2025quantumvacuumenergyorigin,
      title={Quantum Vacuum energy as the origin of Gravity}, 
      author={André LeClair},
      year={2025},
      eprint={2509.02636},
      archivePrefix={arXiv},
      primaryClass={hep-th},
      url={https://arxiv.org/abs/2509.02636}, 
}

@article{DESI:2025zgx,
    author = "Abdul Karim, M. and others",
    collaboration = "DESI",
    title = "{DESI DR2 Results II: Measurements of Baryon Acoustic Oscillations and Cosmological Constraints}",
    eprint = "2503.14738",
    archivePrefix = "arXiv",
    primaryClass = "astro-ph.CO",
    reportNumber = "FERMILAB-PUB-25-0169-PPD",
    month = "3",
    year = "2025"
}

@article{CPL1,
  title="{Accelerating universes with scaling dark matter}",
  author={Chevallier, Michel and Polarski, David},
  journal={International Journal of Modern Physics D},
  volume={10},
  number={02},
  pages={213--223},
  year={2001},
  publisher={World Scientific}
}

@article{CPL2,
  title="{Exploring the expansion history of the universe}",
  author={Linder, Eric V},
  journal={Physical review letters},
  volume={90},
  number={9},
  pages={091301},
  year={2003},
  publisher={APS}
}

@article{Planck2018,
  title="{Planck 2018 results-VI. Cosmological parameters}",
  author={Aghanim, Nabila and Akrami, Yashar and Ashdown, Mark and Aumont, Jonathan and Baccigalupi, Carlo and Ballardini, Mario and Banday, Anthony J and Barreiro, RB and Bartolo, Nicola and Basak, S and others},
  journal={Astronomy \& Astrophysics},
  volume={641},
  pages={A6},
  year={2020},
  publisher={EDP sciences}
}

@article{DiValentino:2026uua,
    author = "Di Valentino, Eleonora",
    title = "{Cracks in the Standard Cosmological Model: Anomalies, Tensions, and Hints of New Physics}",
    eprint = "2601.01525",
    archivePrefix = "arXiv",
    primaryClass = "astro-ph.CO",
    doi = "10.22323/1.507.0012",
    journal = "PoS",
    volume = "COSMICWISPers2025",
    pages = "012",
    year = "2026"
}

@article{Giare:2025ath,
    author = "Giar{\`e}, William and Mena, Olga and Specogna, Enrico and Di Valentino, Eleonora",
    title = "{Neutrino mass tension or suppressed growth rate of matter perturbations?}",
    eprint = "2507.01848",
    archivePrefix = "arXiv",
    primaryClass = "astro-ph.CO",
    doi = "10.1103/njfc-pd1w",
    journal = "Phys. Rev. D",
    volume = "112",
    number = "10",
    pages = "103520",
    year = "2025"
}

@article{SH0ES,
  title="{A comprehensive measurement of the local value of the Hubble constant with 1 km s$^{-1}$ Mpc${-1}$ uncertainty from the Hubble Space Telescope and the SH0ES team}",
  author={Riess, Adam G and Yuan, Wenlong and Macri, Lucas M and Scolnic, Dan and Brout, Dillon and Casertano, Stefano and Jones, David O and Murakami, Yukei and Anand, Gagandeep S and Breuval, Louise and others},
  journal={The Astrophysical journal letters},
  volume={934},
  number={1},
  pages={L7},
  year={2022},
  DOI={10.3847/2041-8213/ac5c5b},
  publisher={IOP Publishing}
}

@article{DOnofrio:2025cuk,
    author = "D'Onofrio, Simone and Odintsov, Sergei and Schiavone, Tiziano",
    title = "{Exponential $f(R)$ cosmology with massive neutrinos as a dynamical dark energy framework}",
    eprint = "2511.06924",
    archivePrefix = "arXiv",
    primaryClass = "gr-qc",
    month = "11",
    year = "2025"
}

@article{Schiavone:2026agq,
    author = "Schiavone, Tiziano and De Angelis, Mariaveronica and Escamilla, Luis A. and Montani, Giovanni and Di Valentino, Eleonora",
    title = "{Revisiting the Matter Creation Process: Observational Constraints on Gravitationally Induced Dark Energy and the Hubble Tension}",
    eprint = "2601.14222",
    archivePrefix = "arXiv",
    primaryClass = "astro-ph.CO",
    month = "1",
    year = "2026"
}

@book{CQG,
author = {Cianfrani, Francesco and Lecian, Orchidea Maria and Lulli, Matteo and Montani, Giovanni},
year = {2014},
month = {07},
pages = {},
title = "{Canonical Quantum Gravity: Fundamentals and Recent Developments}",
isbn = {978-981-4556-64-4},
doi = {10.1142/8957}
}

@article{efstathiou_planck,
  title={The evidence for a spatially flat Universe},
  author={Efstathiou, George and Gratton, Steven},
  journal={Monthly Notices of the Royal Astronomical Society: Letters},
  volume={496},
  number={1},
  pages={L91--L95},
  year={2020},
  publisher={Oxford University Press}
}

@article{Montani_2017,
   title={Bianchi I cosmology in the presence of a causally regularized viscous fluid},
   volume={77},
   ISSN={1434-6052},
   url={http://dx.doi.org/10.1140/epjc/s10052-017-5042-z},
   DOI={10.1140/epjc/s10052-017-5042-z},
   number={7},
   journal={The European Physical Journal C},
   publisher={Springer Science and Business Media LLC},
   author={Montani, Giovanni and Venanzi, Marta},
   year={2017},
   month=jul }

@article{CosmoVerseNetwork:2025alb,
    author = "Di Valentino, Eleonora and others",
    collaboration = "CosmoVerse Network",
    title = "{The CosmoVerse White Paper: Addressing observational tensions in cosmology with systematics and fundamental physics}",
    eprint = "2504.01669",
    archivePrefix = "arXiv",
    primaryClass = "astro-ph.CO",
    doi = "10.1016/j.dark.2025.101965",
    journal = "Phys. Dark Univ.",
    volume = "49",
    pages = "101965",
    year = "2025"
}

@article{divalentino-Hubbletension,
doi = {10.1088/1361-6382/ac086d},
year = {2021},
month = {7},
publisher = {IOP Publishing},
volume = {38},
number = {15},
pages = {153001},
author = {Di Valentino, Eleonora and Mena, Olga and Supriya Pan and Luca Visinelli and Weiqiang Yang and Alessandro Melchiorri and David F Mota and Adam G Riess and Joseph Silk},
title = "{In the realm of the Hubble tension—a review of solutions}",
journal = {Classical and Quantum Gravity},
}

@article{Linder:2024rdj,
    author = "Linder, Eric V.",
    title = "{Interpreting Dark Energy Data Away from $\Lambda$}",
    eprint = "2410.10981",
    archivePrefix = "arXiv",
    primaryClass = "astro-ph.CO",
    month = "10",
    year = "2024"
}

@article{Tsujikawa:2010DE,
    author = "Tsujikawa, Shinji",
    title = "{Dark energy: investigation and modeling}",
    eprint = "1004.1493",
    archivePrefix = "arXiv",
    primaryClass = "astro-ph.CO",
    doi = "10.1007/978-90-481-8685-3_8",
    month = "4",
    year = "2010"
}

@article{Li:2011DE,
    author = "Li, Miao and Li, Xiao-Dong and Wang, Shuang and Wang, Yi",
    title = "{Dark Energy}",
    eprint = "1103.5870",
    archivePrefix = "arXiv",
    primaryClass = "astro-ph.CO",
    reportNumber = "CAS-KITPC-ITP-249",
    doi = "10.1088/0253-6102/56/3/24",
    journal = "Commun. Theor. Phys.",
    volume = "56",
    pages = "525--604",
    year = "2011"
}

@article{Giare_Robust_DDE,
   title={Robust preference for Dynamical Dark Energy in DESI BAO and SN measurements},
   volume={2024},
   ISSN={1475-7516},
   url={http://dx.doi.org/10.1088/1475-7516/2024/10/035},
   DOI={10.1088/1475-7516/2024/10/035},
   number={10},
   journal={Journal of Cosmology and Astroparticle Physics},
   publisher={IOP Publishing},
   author={Giar{\`e}, William and Najafi, Mahdi and Pan, Supriya and Di Valentino, Eleonora and Firouzjaee, Javad T.},
   year={2024},
   month=oct, pages={035} }

@article{giare_dynamical,
  title={Dynamical Dark Energy Beyond Planck? Constraints from multiple CMB probes, DESI BAO and Type-Ia Supernovae},
  author={Giar{\`e}, William},
  journal={arXiv preprint arXiv:2409.17074},
  year={2024}
}

@article{giare_overviewDDE,
    author = "Giar{\`e}, William and Mahassen, Tariq and Di Valentino, Eleonora and Pan, Supriya",
    title = "{An overview of what current data can (and cannot yet) say about evolving dark energy}",
    eprint = "2502.10264",
    archivePrefix = "arXiv",
    primaryClass = "astro-ph.CO",
    doi = "10.1016/j.dark.2025.101906",
    journal = "Phys. Dark Univ.",
    volume = "48",
    pages = "101906",
    year = "2025"
}

@article{giare_interacting,
    author = "Giar{\`e}, William and Sabogal, Miguel A. and Nunes, Rafael C. and Di Valentino, Eleonora",
    title = "{Interacting Dark Energy after DESI Baryon Acoustic Oscillation Measurements}",
    eprint = "2404.15232",
    archivePrefix = "arXiv",
    primaryClass = "astro-ph.CO",
    doi = "10.1103/PhysRevLett.133.251003",
    journal = "Phys. Rev. Lett.",
    volume = "133",
    number = "25",
    pages = "251003",
    year = "2024"
}

@article{divalentino_interacting,
    author = "Silva, Emanuelly and Sabogal, Miguel A. and Scherer, Mateus and Nunes, Rafael C. and Di Valentino, Eleonora and Kumar, Suresh",
    title = "{New constraints on interacting dark energy from DESI DR2 BAO observations}",
    eprint = "2503.23225",
    archivePrefix = "arXiv",
    primaryClass = "astro-ph.CO",
    doi = "10.1103/qqc6-76z4",
    journal = "Phys. Rev. D",
    volume = "111",
    number = "12",
    pages = "123511",
    year = "2025"
}

@article{Zhai:2025hfi,
    author = "Zhai, Yuejia and de Cesare, Marco and van de Bruck, Carsten and Di Valentino, Eleonora and Wilson-Ewing, Edward",
    title = "{A low-redshift preference for an interacting dark energy model}",
    eprint = "2503.15659",
    archivePrefix = "arXiv",
    primaryClass = "astro-ph.CO",
    month = "3",
    year = "2025"
}

@article{Kessler:2025kju,
    author = "Kessler, Daniel A. and Escamilla, Luis A. and Pan, Supriya and Di Valentino, Eleonora",
    title = "{One-parameter dynamical dark energy: Hints for oscillations}",
    eprint = "2504.00776",
    archivePrefix = "arXiv",
    primaryClass = "astro-ph.CO",
    month = "4",
    year = "2025"
}

@article{Fazzari:2025lzd,
    author = "Fazzari, Elisa and Giar{\`e}, William and Di Valentino, Eleonora",
    title = "{Cosmographic Footprints of Dynamical Dark Energy}",
    eprint = "2509.16196",
    archivePrefix = "arXiv",
    primaryClass = "astro-ph.CO",
    month = "9",
    year = "2025"
}

@article{DESI:2025fii,
    author = "Lodha, K. and others",
    collaboration = "DESI",
    title = "{Extended Dark Energy analysis using DESI DR2 BAO measurements}",
    eprint = "2503.14743",
    archivePrefix = "arXiv",
    primaryClass = "astro-ph.CO",
    reportNumber = "FERMILAB-PUB-25-0164-PPD",
    month = "3",
    year = "2025"
}

@article{Gonzalez-Fuentes:2025lei,
    author = "Gonz{\'a}lez-Fuentes, Alex and G{\'o}mez-Valent, Adri{\`a}",
    title = "{Reconstruction of dark energy and late-time cosmic expansion using the Weighted Function Regression method}",
    eprint = "2506.11758",
    archivePrefix = "arXiv",
    primaryClass = "astro-ph.CO",
    month = "6",
    year = "2025"
}

@article{Berti:2025phi,
    author = "Berti, Maria and Bellini, Emilio and Bonvin, Camille and Kunz, Martin and Viel, Matteo and Zumalacarregui, Miguel",
    title = "{Reconstructing the dark energy density in light of DESI BAO observations}",
    eprint = "2503.13198",
    archivePrefix = "arXiv",
    primaryClass = "astro-ph.CO",
    doi = "10.1103/dj3k-84v4",
    journal = "Phys. Rev. D",
    volume = "112",
    number = "2",
    pages = "023518",
    year = "2025"
}

@article{Nesseris:2012tt,
    author = "Nesseris, Savvas and Garcia-Bellido, Juan",
    title = "{A new perspective on Dark Energy modeling via Genetic Algorithms}",
    eprint = "1205.0364",
    archivePrefix = "arXiv",
    primaryClass = "astro-ph.CO",
    reportNumber = "IFT-UAM-CSIC-12-42",
    doi = "10.1088/1475-7516/2012/11/033",
    journal = "JCAP",
    volume = "11",
    pages = "033",
    year = "2012"
}

@article{schiavone2024,
    author = "Schiavone, Tiziano and Montani, Giovanni",
    title = "{Evolution of an effective Hubble constant in f (R) modified gravity}",
    eprint = "2408.01410",
    archivePrefix = "arXiv",
    primaryClass = "gr-qc",
    doi = "10.1393/ncc/i2025-25105-3",
    journal = "Nuovo Cim. C",
    volume = "48",
    number = "3",
    pages = "105",
    year = "2025"
}

@article{kazantzidis2020,
  title="{Hints of a local matter underdensity or modified gravity in the low z Pantheon data}",
  author={Kazantzidis, L and Perivolaropoulos, L},
  journal={Physical Review D},
  volume={102},
  number={2},
  pages={023520},
  year={2020},
  publisher={APS}
}

@article{dainotti2021,
  title="{On the Hubble constant tension in the SNe Ia Pantheon sample}",
  author={Dainotti, Maria Giovanna and De Simone, Biagio and Schiavone, Tiziano and Montani, Giovanni and Rinaldi, Enrico and Lambiase, Gaetano},
  journal={The Astrophysical Journal},
  volume={912},
  number={2},
  pages={150},
  year={2021},
  publisher={IOP Publishing}
}

@article{dainotti2022,
  title="{On the evolution of the Hubble constant with the SNe Ia pantheon sample and baryon acoustic oscillations: a feasibility study for GRB-cosmology in 2030}",
  author={Dainotti, Maria Giovanna and De Simone, Biagio and Schiavone, Tiziano and Montani, Giovanni and Rinaldi, Enrico and Lambiase, Gaetano and Bogdan, Malgorzata and Ugale, Sahil},
  journal={Galaxies},
  volume={10},
  number={1},
  pages={24},
  year={2022},
  publisher={MDPI}
}

@article{schiavone2023,
   title="{f(R) gravity in the Jordan frame as a paradigm for the Hubble tension}",
   volume={522},
   DOI={10.1093/mnrasl/slad041},
   number={1},
   journal={Monthly Notices of the Royal Astronomical Society: Letters},
   publisher={Oxford University Press (OUP)},
   author={Schiavone, Tiziano and Montani, Giovanni and Bombacigno, Flavio},
   year={2023},
   month=mar, pages={L72–L77} 
}

@article{Montani_carlevaro_dainotti,
    author = "Montani, Giovanni and Carlevaro, Nakia and Dainotti, Maria G.",
    title = "{Running Hubble constant: Evolutionary Dark Energy}",
    eprint = "2411.07060",
    archivePrefix = "arXiv",
    primaryClass = "gr-qc",
    doi = "10.1016/j.dark.2025.101847",
    journal = "Phys. Dark Univ.",
    volume = "48",
    pages = "101847",
    year = "2025"
}

@article{Montani_deangelis_dainotti,
    author = "Montani, Giovanni and De Angelis, Mariaveronica and Dainotti, Maria Giovanna",
    title = "{Decay of dark energy into dark matter in a metric $f(R)$ gravity: Effective running Hubble constant}",
    eprint = "2506.13288",
    archivePrefix = "arXiv",
    primaryClass = "astro-ph.CO",
    doi = "10.1016/j.dark.2025.101969",
    journal = "Phys. Dark Univ.",
    volume = "49",
    pages = "101969",
    year = "2025"
}

@article{Montani_deangelis_fR,
    author = "Montani, Giovanni and De Angelis, Mariaveronica and Bombacigno, Flavio and Carlevaro, Nakia",
    title = "{Metric f(R) gravity with dynamical dark energy as a scenario for the Hubble tension}",
    eprint = "2306.11101",
    archivePrefix = "arXiv",
    primaryClass = "gr-qc",
    doi = "10.1093/mnrasl/slad159",
    journal = "Mon. Not. Roy. Astron. Soc.",
    volume = "527",
    number = "1",
    pages = "L156--L161",
    year = "2023"
}

@article{colgain_Dainotti,
  title="{Does DESI 2024 Confirm $\Lambda$CDM?}",
  author={Colg{\'a}in, Eoin {\'O} and Dainotti, Maria Giovanna and Capozziello, Salvatore and Pourojaghi, Saeed and Sheikh-Jabbari, MM and Stojkovic, Dejan},
  journal={arXiv preprint arXiv:2404.08633},
  year={2024}
}

@article{elizalde_odintsov,
  title="{Can we learn from matter creation to solve the $H_0$ tension problem?}",
  author={Elizalde, Emilio and Khurshudyan, Martiros and Odintsov, Sergei D},
  journal={The European Physical Journal C},
  volume={84},
  number={8},
  pages={782},
  year={2024},
  publisher={Springer}
}

@article{matcre_montani2001,
  title="{Influence of particle creation on flat and negative curved FLRW universes}",
  author={Montani, Giovanni},
  journal={Classical and Quantum Gravity},
  volume={18},
  number={1},
  pages={193},
  year={2001},
  publisher={IOP Publishing}
}

@article{matcre_calvaoLima,
  title="{On the thermodynamics of matter creation in cosmology}",
  author={Calvao, MO and Lima, JAS and Waga, I},
  journal={Physics Letters A},
  volume={162},
  number={3},
  pages={223--226},
  year={1992},
  publisher={Elsevier}
}

@article{matcre_nunes2015,
  title="{Phantom behavior via cosmological creation of particles}",
  author={Nunes, Rafael C and Pav{\'o}n, Diego},
  journal={Physical Review D},
  volume={91},
  number={6},
  pages={063526},
  year={2015},
  publisher={APS}
}

@article{Belinskii1975,
    author = {Belinskii, V. A. and Khalatnikov, I. M.},
    title = {Influence of viscosity on the character of cosmological evolution},
    journal = {Soviet JETT},
    volume = {42},
    pages = {205},
    year = {1975}
}

@article{Belinskii1979,
    author = {Belinskii, V. A. and Nikomarov, E. S. and Khalatnikov, I. M.},
    title = {Investigation of the cosmological evolution of viscoelastic matter with causal thermodynamics},
    journal = {Soviet JETT},
    volume = {50},
    number = {2},
    pages = {21},
    year = {1979}
}

@article{Belinskii1977,
    author = {Belinskii, V. A. and Khalatnikov, I. M.},
    title = {Viscosity effects in isotropic cosmologies},
    journal = {Soviet JETT},
    volume = {45},
    pages = {19},
    year = {1977}
}

@article{Freaza:2002ic,
    author = "Freaza, M. P. and de Souza, R. S. and Waga, I.",
    title = "{Cosmic acceleration and matter creation}",
    doi = "10.1103/PhysRevD.66.103502",
    journal = "Phys. Rev. D",
    volume = "66",
    pages = "103502",
    year = "2002"
}

@article{Cardenas:2020grl,
    author = "C{\'a}rdenas, V{\'\i}ctor H. and Cruz, Miguel and Lepe, Samuel and Nojiri, Shin'ichi and Odintsov, Sergei D.",
    title = "{Challenging matter creation models in the phantom divide}",
    eprint = "2004.02337",
    archivePrefix = "arXiv",
    primaryClass = "gr-qc",
    doi = "10.1103/PhysRevD.101.083530",
    journal = "Phys. Rev. D",
    volume = "101",
    number = "8",
    pages = "083530",
    year = "2020"
}

@article{Fazzari:2025mww,
    author = "Fazzari, E. and Dainotti, M. G. and Montani, G. and Melchiorri, A.",
    title = "{The effective running Hubble constant in SNe Ia as a marker for the dark energy nature}",
    eprint = "2506.04162",
    archivePrefix = "arXiv",
    primaryClass = "astro-ph.CO",
    doi = "10.1016/j.jheap.2025.100459",
    journal = "JHEAp",
    volume = "49",
    pages = "100459",
    year = "2026",
}

@article{Montani:2025rcy,
    author = "Montani, Giovanni and Fazzari, Elisa and Carlevaro, Nakia and Dainotti, Maria Giovanna",
    title = "{Two Dynamical Scenarios for Binned Master Sample Interpretation}",
    eprint = "2507.14048",
    archivePrefix = "arXiv",
    primaryClass = "astro-ph.CO",
    doi = "10.3390/e27090895",
    journal = "Entropy",
    volume = "27",
    pages = "895",
    year = "2025"
}

@article{Bayes_trotta,
  title="{Bayes in the sky: Bayesian inference and model selection in cosmology}",
  author={Trotta, Roberto},
  journal={Contemporary Physics},
  volume={49},
  number={2},
  pages={71--104},
  year={2008},
  publisher={Taylor \& Francis}
}

@book{jeffreys_scale,
  title={The theory of probability},
  author={Jeffreys, Harold},
  year={1998},
  publisher={OuP Oxford}
}

@article{wagenmakers2007,
author = {Wagenmakers, Eric-Jan},
year = {2007},
month = {11},
pages = {779-804},
title = "{A Practical Solution to the Pervasive Problems of p Values}",
volume = {14},
journal = {Psychonomic bulletin \& review},
doi = {10.3758/BF03194105}
}

@article{cobaya,
  title="{Cobaya: Code for Bayesian Analysis of hierarchical physical models}",
  author={Torrado, Jesus and Lewis, Antony},
  journal={Journal of Cosmology and Astroparticle Physics},
  volume={2021},
  number={05},
  pages={057},
  year={2021},
  publisher={IOP Publishing}
}

@article{getdist,
   author = "Lewis, Antony",
   title = "{GetDist: a Python package for analysing Monte Carlo samples}",
   eprint = "1910.13970",
   archivePrefix = "arXiv",
   primaryClass = "astro-ph.IM",
   doi = "10.1088/1475-7516/2025/08/025",
   journal = "JCAP",
   volume = "08",
   pages = "025",
   year = "2025"
}

@article{gelman_rubin,
  title="{Inference from iterative simulation using multiple sequences}",
  author={Gelman, Andrew and Rubin, Donald B},
  journal={Statistical science},
  volume={7},
  number={4},
  pages={457--472},
  year={1992},
  publisher={Institute of Mathematical Statistics}
}

@article{Dainotti:2025qxz,
    author = "Dainotti, M. G. and others",
    title = "{A New Master Supernovae Ia sample and the investigation of the Hubble tension}",
    eprint = "2501.11772",
    archivePrefix = "arXiv",
    primaryClass = "astro-ph.CO",
    reportNumber = "KEK-TH-2711, KEK-Cosmo-0378",
    doi = "10.1016/j.jheap.2025.100405",
    journal = "JHEAp",
    volume = "48",
    pages = "100405",
    year = "2025"
}

@article{JLA,
  title="{Improved cosmological constraints from a joint analysis of the SDSS-II and SNLS supernova samples}",
  author={Betoule, Marc and Kessler, R and Guy, J and Mosher, J and Hardin, D and Biswas, R and Astier, P and El-Hage, P and Konig, M and Kuhlmann, S and others},
  journal={Astronomy \& Astrophysics},
  volume={568},
  pages={A22},
  year={2014},
  publisher={EDP Sciences}
}

@article{DES:2024jxu,
    author = "Abbott, T. M. C. and others",
    collaboration = "DES",
    title = "{The Dark Energy Survey: Cosmology Results with {\ensuremath{\sim}}1500 New High-redshift Type Ia Supernovae Using the Full 5 yr Data Set}",
    eprint = "2401.02929",
    archivePrefix = "arXiv",
    primaryClass = "astro-ph.CO",
    reportNumber = "FERMILAB-PUB-23-0821-PPD, DES-2023-805",
    doi = "10.3847/2041-8213/ad6f9f",
    journal = "Astrophys. J. Lett.",
    volume = "973",
    number = "1",
    pages = "L14",
    year = "2024"
}

@article{scolnic2018,
  title="{The complete light-curve sample of spectroscopically confirmed SNe Ia from Pan-STARRS1 and cosmological constraints from the combined pantheon sample}",
  author={Scolnic, Daniel Moshe and Jones, DO and Rest, A and Pan, YC and Chornock, R and Foley, RJ and Huber, ME and Kessler, R and Narayan, Gautham and Riess, AG and others},
  journal={The Astrophysical Journal},
  volume={859},
  number={2},
  pages={101},
  year={2018},
  publisher={IOP Publishing}
}

@article{Scolnic:2021amr,
    author = "Scolnic, Dan and others",
    title = "{The Pantheon+ Analysis: The Full Data Set and Light-curve Release}",
    eprint = "2112.03863",
    archivePrefix = "arXiv",
    primaryClass = "astro-ph.CO",
    doi = "10.3847/1538-4357/ac8b7a",
    journal = "Astrophys. J.",
    volume = "938",
    number = "2",
    pages = "113",
    year = "2022"
}

@article{Brout:2022vxf,
    author = "Brout, Dillon and others",
    title = "{The Pantheon+ Analysis: Cosmological Constraints}",
    eprint = "2202.04077",
    archivePrefix = "arXiv",
    primaryClass = "astro-ph.CO",
    doi = "10.3847/1538-4357/ac8e04",
    journal = "Astrophys. J.",
    volume = "938",
    number = "2",
    pages = "110",
    year = "2022"
}

@article{CC_borghi,
  title="{Toward a better understanding of cosmic chronometers: a new measurement of $H(z)$ at $z\sim0.7$}",
  author={Borghi, Nicola and Moresco, Michele and Cimatti, Andrea},
  journal={The Astrophysical Journal Letters},
  volume={928},
  number={1},
  pages={L4},
  year={2022},
  publisher={IOP Publishing}
}

@article{Jimenez_247,
   title="{Constraining Cosmological Parameters Based on Relative Galaxy Ages}",
   volume={573},
   DOI={10.1086/340549},
   number={1},
   journal={The Astrophysical Journal},
   publisher={American Astronomical Society},
   author={Jimenez, Raul and Loeb, Abraham},
   year={2002},
   month=jul, 
    pages={37–42} 
}

@article{Moresco_2012,
   title="{Improved constraints on the expansion rate of the Universe up to $z \sim 1.1$ from the spectroscopic evolution of cosmic chronometers}",
   volume={2012},
   ISSN={1475-7516},
   url={http://dx.doi.org/10.1088/1475-7516/2012/08/006},
   DOI={10.1088/1475-7516/2012/08/006},
   number={08},
   journal={Journal of Cosmology and Astroparticle Physics},
   publisher={IOP Publishing},
   author={M. Moresco and A. Cimatti and R. Jimenez and others},
   year={2012},
   month=aug, pages={006–006} }

@article{Moresco_2015,
   title="{Raising the bar: new constraints on the Hubble parameter with cosmic chronometers at $z\sim2$}",
   volume={450},
   ISSN={1745-3925},
   url={http://dx.doi.org/10.1093/mnrasl/slv037},
   DOI={10.1093/mnrasl/slv037},
   number={1},
   journal={Monthly Notices of the Royal Astronomical Society: Letters},
   publisher={Oxford University Press (OUP)},
   author={Moresco, Michele},
   year={2015},
   month=apr, pages={L16–L20} }

@article{Moresco_2016,
   title="{A $6\%$ measurement of the Hubble parameter at $z\sim0.45$: direct evidence of the epoch of cosmic re-acceleration}",
   volume={2016},
   ISSN={1475-7516},
   url={http://dx.doi.org/10.1088/1475-7516/2016/05/014},
   DOI={10.1088/1475-7516/2016/05/014},
   number={05},
   journal={Journal of Cosmology and Astroparticle Physics},
   publisher={IOP Publishing},
   author={Moresco, Michele and Pozzetti, Lucia and Cimatti, Andrea and Jimenez, Raul and Maraston, Claudia and Verde, Licia and Thomas, Daniel and Citro, Annalisa and Tojeiro, Rita and Wilkinson, David},
   year={2016},
   month=may, pages={014–014} }

@article{Yarahmadi:2025fml,
    author = "Yarahmadi, Muhammad and Salehi, Amin",
    title = "{A comparative Bayesian PINN{\textendash}MCMC analysis of Barrow{\textendash}Tsallis holographic dark energy with neutrinos: Toward resolving the Hubble tension}",
    doi = "10.1016/j.jheap.2025.100498",
    journal = "JHEAp",
    volume = "50",
    pages = "100498",
    year = "2026"
}

@article{Wang:2025xvi,
    author = "Wang, Yi-Ying and Li, Yin-Jie and Fan, Yi-Zhong",
    title = "{Evidence for the dynamical dark energy with evolving Hubble constant}",
    eprint = "2510.14390",
    archivePrefix = "arXiv",
    primaryClass = "astro-ph.CO",
    month = "10",
    year = "2025"
}

@article{Yashiki:2025loj,
    author = "Yashiki, Mai",
    title = "{Toward a simultaneous resolution of the H0 and S8 tensions: Early dark energy and an interacting dark sector model}",
    eprint = "2505.23382",
    archivePrefix = "arXiv",
    primaryClass = "astro-ph.CO",
    doi = "10.1103/qw1d-mdrz",
    journal = "Phys. Rev. D",
    volume = "112",
    number = "6",
    pages = "063517",
    year = "2025"
}

@misc{manoharan2025solveshubbletensionphenomenological,
      title={What solves the Hubble tension in phenomenological dark energy models at background level?}, 
      author={Manosh T. Manoharan},
      year={2025},
      eprint={2505.24743},
      archivePrefix={arXiv},
      primaryClass={astro-ph.CO},
      url={https://arxiv.org/abs/2505.24743}, 
}

@misc{wang2025universeexperienceadslandscape,
      title={Can the universe experience an AdS landscape since matter-radiation equality?}, 
      author={Hao Wang and Yun-Song Piao},
      year={2025},
      eprint={2506.04306},
      archivePrefix={arXiv},
      primaryClass={gr-qc},
      url={https://arxiv.org/abs/2506.04306}, 
}

@misc{goswami2025constrainingfrlmgravity,
      title={Constraining a $f(R, L_m)$ Gravity Cosmological Model with Observational Data}, 
      author={G. K. Goswami and Anirudh Pradhan},
      year={2025},
      eprint={2505.18226},
      archivePrefix={arXiv},
      primaryClass={gr-qc},
      url={https://arxiv.org/abs/2505.18226}, 
}

@misc{carloni2025addressingh0tensionmatter,
      title={Addressing the $H_0$ tension through matter with pressure and no early dark energy}, 
      author={Youri Carloni and Orlando Luongo and Marco Muccino},
      year={2025},
      eprint={2506.11531},
      archivePrefix={arXiv},
      primaryClass={astro-ph.CO},
      url={https://arxiv.org/abs/2506.11531}, 
}

@misc{kalita2025revealinglimitationstandardcosmological,
      title={Revealing Limitation in the Standard Cosmological Model: A Redshift-Dependent Hubble Constant from Fast Radio Bursts}, 
      author={Surajit Kalita and Akhil Uniyal and Tomasz Bulik and Yosuke Mizuno},
      year={2025},
      eprint={2506.14947},
      archivePrefix={arXiv},
      primaryClass={astro-ph.CO},
      url={https://arxiv.org/abs/2506.14947}, 
}

@misc{lee2025alleviatinghubbletensioncosmological,
      title={Alleviating the Hubble Tension via Cosmological Time Dilation in the meVSL Model}, 
      author={Seokcheon Lee},
      year={2025},
      eprint={2509.08840},
      archivePrefix={arXiv},
      primaryClass={physics.gen-ph},
      url={https://arxiv.org/abs/2509.08840}, 
}

@article{Dixit_2025,
   title={Beyond $\Lambda$ CDM: Exploring a dynamical cosmological constant framework consistent with late-time observations},
   volume={483},
   ISSN={0003-4916},
   url={http://dx.doi.org/10.1016/j.aop.2025.170275},
   DOI={10.1016/j.aop.2025.170275},
   journal={Annals of Physics},
   publisher={Elsevier BV},
   author={Dixit, Archana and Yadav, Manish and Pradhan, Anirudh and Barak, M.S.},
   year={2025},
   month=dec, pages={170275} }

@misc{du2025modelindependentlateuniversemeasurementsh0,
      title={Model-independent late-universe measurements of $H_0$ and $\Omega_\mathrm{K}$ with the PAge-improved inverse distance ladder}, 
      author={Guo-Hong Du and Tian-Nuo Li and Jia-Le Ling and Yan-Hong Yao and Jing-Fei Zhang and Xin Zhang},
      year={2025},
      eprint={2510.26355},
      archivePrefix={arXiv},
      primaryClass={astro-ph.CO},
      url={https://arxiv.org/abs/2510.26355}, 
}

@article{articleGurzadyan,
author = {Gurzadyan, V. and Fimin, Nikolay and Chechetkin, V.M.},
year = {2025},
month = {01},
pages = {},
title = {Cosmic voids and the kinetic analysis IV. Hubble tension and the cosmological constant},
volume = {694},
journal = {Astronomy {\&} Astrophysics},
doi = {10.1051/0004-6361/202553679}
}

@misc{efstratiou2025addressingdesidr2phantomcrossing,
      title={Addressing the DESI DR2 Phantom-Crossing Anomaly and Enhanced $H_0$ Tension with Reconstructed Scalar-Tensor Gravity}, 
      author={Dimitrios Efstratiou and Evangelos Achilleas Paraskevas and Leandros Perivolaropoulos},
      year={2025},
      eprint={2511.04610},
      archivePrefix={arXiv},
      primaryClass={astro-ph.CO},
      url={https://arxiv.org/abs/2511.04610}, 
}

@misc{lee2025geometricinterpretationredshiftevolution,
      title={Geometric Interpretation of the Redshift Evolution of $h_0(z)$}, 
      author={Seokcheon Lee},
      year={2025},
      eprint={2511.07454},
      archivePrefix={arXiv},
      primaryClass={physics.gen-ph},
      url={https://arxiv.org/abs/2511.07454}, 
}

@ARTICLE{grb2014ApJ...783..126P,
       author = {{Postnikov}, S. and {Dainotti}, M.~G. and {Hernandez}, X. and {Capozziello}, S.},
        title = "{Nonparametric Study of the Evolution of the Cosmological Equation of State with SNeIa, BAO, and High-redshift GRBs}",
      journal = {\apj},
         year = 2014,
        month = mar,
       volume = {783},
       number = {2},
          eid = {126},
        pages = {126},
          doi = {10.1088/0004-637X/783/2/126},
archivePrefix = {arXiv},
       eprint = {1401.2939},
 primaryClass = {astro-ph.CO},
       adsurl = {https://ui.adsabs.harvard.edu/abs/2014ApJ...783..126P}
}

@ARTICLE{grb2024JCAP...08..015A,
       author = {{Adil}, Shahnawaz A. and {Dainotti}, Maria G. and {Sen}, Anjan A.},
        title = "{Revisiting the concordance {\ensuremath{\Lambda}}CDM model using Gamma-Ray Bursts together with supernovae Ia and Planck data}",
      journal = {\jcap},
         year = 2024,
        month = aug,
       volume = {2024},
       number = {8},
          eid = {015},
        pages = {015},
          doi = {10.1088/1475-7516/2024/08/015},
archivePrefix = {arXiv},
       eprint = {2405.01452},
 primaryClass = {astro-ph.HE},
       adsurl = {https://ui.adsabs.harvard.edu/abs/2024JCAP...08..015A}
}

@ARTICLE{grb2022PASJ...74.1095D,
       author = {{Dainotti}, Maria Giovanna and {Sarracino}, Giuseppe and {Capozziello}, Salvatore},
        title = "{Gamma-ray bursts, supernovae Ia, and baryon acoustic oscillations: A binned cosmological analysis}",
      journal = {\pasj},
         year = 2022,
        month = oct,
       volume = {74},
       number = {5},
        pages = {1095-1113},
          doi = {10.1093/pasj/psac057},
archivePrefix = {arXiv},
       eprint = {2206.07479},
 primaryClass = {astro-ph.CO},
       adsurl = {https://ui.adsabs.harvard.edu/abs/2022PASJ...74.1095D}
}

@ARTICLE{qua2022ApJ...931..106D,
       author = {{Dainotti}, Maria Giovanna and {Bargiacchi}, Giada and {Lenart}, Aleksander {\L}ukasz and {Capozziello}, Salvatore and {{\'O} Colg{\'a}in}, Eoin and {Solomon}, Rance and {Stojkovic}, Dejan and {Sheikh-Jabbari}, M.~M.},
        title = "{Quasar Standardization: Overcoming Selection Biases and Redshift Evolution}",
      journal = {\apj},
         year = 2022,
        month = jun,
       volume = {931},
       number = {2},
          eid = {106},
        pages = {106},
          doi = {10.3847/1538-4357/ac6593},
archivePrefix = {arXiv},
       eprint = {2203.12914},
 primaryClass = {astro-ph.HE},
       adsurl = {https://ui.adsabs.harvard.edu/abs/2022ApJ...931..106D}
}

@ARTICLE{qua2023ApJ...950...45D,
       author = {{Dainotti}, M.~G. and {Bargiacchi}, G. and {Lenart}, A. {\L}. and {Nagataki}, S. and {Capozziello}, S.},
        title = "{Quasars: Standard Candles up to z = 7.5 with the Precision of Supernovae Ia}",
      journal = {\apj},
         year = 2023,
        month = jun,
       volume = {950},
       number = {1},
          eid = {45},
        pages = {45},
          doi = {10.3847/1538-4357/accea0},
archivePrefix = {arXiv},
       eprint = {2305.19668},
 primaryClass = {astro-ph.CO},
       adsurl = {https://ui.adsabs.harvard.edu/abs/2023ApJ...950...45D}
}

@ARTICLE{qua2023ApJS..264...46L,
       author = {{Lenart}, Aleksander {\L}ukasz and {Bargiacchi}, Giada and {Dainotti}, Maria Giovanna and {Nagataki}, Shigehiro and {Capozziello}, Salvatore},
        title = "{A Bias-free Cosmological Analysis with Quasars Alleviating H $_{0}$ Tension}",
      journal = {\apjs},
         year = 2023,
        month = feb,
       volume = {264},
       number = {2},
          eid = {46},
        pages = {46},
          doi = {10.3847/1538-4365/aca404},
archivePrefix = {arXiv},
       eprint = {2211.10785},
 primaryClass = {astro-ph.CO},
       adsurl = {https://ui.adsabs.harvard.edu/abs/2023ApJS..264...46L}
}

@misc{desimone2024doubletcosmologicalmodelschallenge,
      title={A doublet of cosmological models to challenge the H0 tension in the Pantheon Supernovae Ia catalog}, 
      author={B. De Simone and M. H. P. M. van Putten and M. G. Dainotti and G. Lambiase},
      year={2024},
      eprint={2411.05744},
      archivePrefix={arXiv},
      primaryClass={astro-ph.CO},
      url={https://arxiv.org/abs/2411.05744}, 
}

@misc{su2025exploringjointobservationcsst,
      title={Exploring Joint Observation of the CSST Shear and clustering of astrophysical gravitational wave source measurements}, 
      author={Pengfei Su and Yan Gong and Qi Xiong and Dingao Hu and Hengjie Lin and Furen Deng and Xuelei Chen},
      year={2025},
      eprint={2510.20203},
      archivePrefix={arXiv},
      primaryClass={astro-ph.CO},
      url={https://arxiv.org/abs/2510.20203}, 
}

@misc{pan2025determininghubbleconstantcrosscorrelation,
      title={Determining the Hubble Constant through Cross-Correlation of Galaxies and Gravitational Waves}, 
      author={Jiaming Pan and Dragan Huterer and Camille Avestruz and Damon H. T. Cheung and Emery Trott and Neal Dalal and Donghui Jeong},
      year={2025},
      eprint={2510.19931},
      archivePrefix={arXiv},
      primaryClass={astro-ph.CO},
      url={https://arxiv.org/abs/2510.19931}, 
}

@article{DAINOTTI202430,
title = {On the statistical assumption on the distance moduli of Supernovae Ia and its impact on the determination of cosmological parameters},
journal = {Journal of High Energy Astrophysics},
volume = {41},
pages = {30-41},
year = {2024},
issn = {2214-4048},
doi = {https://doi.org/10.1016/j.jheap.2024.01.001},
url = {https://www.sciencedirect.com/science/article/pii/S2214404824000016},
author = {M.G. Dainotti and G. Bargiacchi and M. Bogdan and S. Capozziello and S. Nagataki},
}

@article{Bargiacchi_2023,
   title={Gamma-ray bursts, quasars, baryonic acoustic oscillations, and supernovae Ia: new statistical insights and cosmological constraints},
   volume={521},
   ISSN={1365-2966},
   url={http://dx.doi.org/10.1093/mnras/stad763},
   DOI={10.1093/mnras/stad763},
   number={3},
   journal={Monthly Notices of the Royal Astronomical Society},
   publisher={Oxford University Press (OUP)},
   author={Bargiacchi, G and Dainotti, M G and Nagataki, S and Capozziello, S},
   year={2023},
   month=mar, pages={3909-3924}, }

@misc{desmond2025subtlestatisticsdistanceladder,
      title={The subtle statistics of the distance ladder: On the distance prior and selection effects}, 
      author={Harry Desmond and Richard Stiskalek and Jose Antonio Najera and Indranil Banik},
      year={2025},
      eprint={2511.03394},
      archivePrefix={arXiv},
      primaryClass={astro-ph.CO},
      url={https://arxiv.org/abs/2511.03394}, 
}

@article{opticalDainotti_2022,
   title={Optical and X-ray GRB Fundamental Planes as cosmological distance indicators},
   volume={514},
   ISSN={1365-2966},
   url={http://dx.doi.org/10.1093/mnras/stac1141},
   DOI={10.1093/mnras/stac1141},
   number={2},
   journal={Monthly Notices of the Royal Astronomical Society},
   publisher={Oxford University Press (OUP)},
   author={Dainotti, M G and Nielson, V and Sarracino, G and Rinaldi, E and Nagataki, S and Capozziello, S and Gnedin, O Y and Bargiacchi, G},
   year={2022},
   month=jun, pages={1828–1856} }

@misc{zhan2025hubbleconstantmeasurementqpes,
      title={Hubble constant measurement from QPEs as electromagnetic counterparts to extreme mass ratio inspirals}, 
      author={Yejing Zhan and Di Wang and Shuang-Xi Yi and Fa-Yin Wang},
      year={2025},
      eprint={2506.14150},
      archivePrefix={arXiv},
      primaryClass={astro-ph.HE},
      url={https://arxiv.org/abs/2506.14150}, 
}

@article{Mukherjee_2025,
   title={Constraining the Hubble parameter with the 21-cm brightness temperature signal in a universe with inhomogeneities},
   volume={112},
   ISSN={2470-0029},
   url={http://dx.doi.org/10.1103/w1wp-tqz2},
   DOI={10.1103/w1wp-tqz2},
   number={6},
   journal={Physical Review D},
   publisher={American Physical Society (APS)},
   author={Mukherjee, Subhadeep and Pandey, Shashank Shekhar and Majumdar, A. S.},
   year={2025},
   month=sep }

@misc{chen2025measuringcosmicdipolegolden,
      title={Measuring the cosmic dipole with golden dark sirens in the era of next-generation ground-based gravitational wave detectors}, 
      author={Anson Chen},
      year={2025},
      eprint={2505.12678},
      archivePrefix={arXiv},
      primaryClass={gr-qc},
      url={https://arxiv.org/abs/2505.12678}, 
}

@article{Disconzi_2015,
   title={New approach to cosmological bulk viscosity},
   volume={91},
   ISSN={1550-2368},
   url={http://dx.doi.org/10.1103/PhysRevD.91.043532},
   DOI={10.1103/physrevd.91.043532},
   number={4},
   journal={Physical Review D},
   publisher={American Physical Society (APS)},
   author={Disconzi, Marcelo M. and Kephart, Thomas W. and Scherrer, Robert J.},
   year={2015},
   month=feb }

@inproceedings{Carlevaro_2008,
   title={Gravitational Stability and Bulk Cosmology},
   volume={966},
   ISSN={0094-243X},
   url={http://dx.doi.org/10.1063/1.2837002},
   DOI={10.1063/1.2837002},
   booktitle={AIP Conference Proceedings},
   publisher={AIP},
   author={Carlevaro, Nakia and Montani, Giovanni and Bianco, Carlo Luciano and Xue, She-Sheng},
   year={2008},
   pages={241-248} }

@book{landau1981physical,
  title        = {Physical Kinetics},
  author       = {Landau, L. D. and Lifshitz, E. M.},
  series       = {Course of Theoretical Physics},
  volume       = {10},
  year         = {1981},
  publisher    = {Pergamon Press},
}

@ARTICLE{2023MNRAS.518.2201D,
       author = {{Dainotti}, M.~G. and {Lenart}, A. {\L}. and {Chraya}, A. and {Sarracino}, G. and {Nagataki}, S. and {Fraija}, N. and {Capozziello}, S. and {Bogdan}, M.},
        title = "{The gamma-ray bursts fundamental plane correlation as a cosmological tool}",
      journal = {\mnras},
         year = 2023,
        month = jan,
       volume = {518},
       number = {2},
        pages = {2201-2240},
          doi = {10.1093/mnras/stac2752},
archivePrefix = {arXiv},
       eprint = {2209.08675},
 primaryClass = {astro-ph.HE},
       adsurl = {https://ui.adsabs.harvard.edu/abs/2023MNRAS.518.2201D}
}

@article{10.1093/mnras/stt1516,
    author = {Dainotti, M. G. and Cardone, V. F. and Piedipalumbo, E. and Capozziello, S.},
    title = {Slope evolution of GRB correlations and cosmology},
    journal = {Monthly Notices of the Royal Astronomical Society},
    volume = {436},
    number = {1},
    pages = {82-88},
    year = {2013},
    month = {09},
    issn = {0035-8711},
    doi = {10.1093/mnras/stt1516},
    url = {https://doi.org/10.1093/mnras/stt1516},
}

@article{Dainotti_2013,
doi = {10.1088/0004-637X/774/2/157},
url = {https://doi.org/10.1088/0004-637X/774/2/157},
year = {2013},
month = {aug},
publisher = {The American Astronomical Society},
volume = {774},
number = {2},
pages = {157},
author = {Dainotti, Maria Giovanna and Petrosian, Vahe' and Singal, Jack and Ostrowski, Michal},
title = {DETERMINATION OF THE INTRINSIC LUMINOSITY TIME CORRELATION IN THE X-RAY AFTERGLOWS OF GAMMA-RAY BURSTS},
journal = {The Astrophysical Journal},
}

@misc{dainotti2024newbinningmethodchoose,
      title={A new binning method to choose a standard set of Quasars}, 
      author={Maria Giovanna Dainotti and Aleksander Lukasz Lenart and Mina Godsi Yengejeh and Satyajit Chakraborty and Nissim Fraija and Eleonora Di Valentino and Giovanni Montani},
      year={2024},
      eprint={2401.12847},
      archivePrefix={arXiv},
      primaryClass={astro-ph.HE},
      url={https://arxiv.org/abs/2401.12847}, 
}

@Article{galaxies12010004,
AUTHOR = {Dainotti, Maria Giovanna and Bargiacchi, Giada and Lenart, Aleksander Łukasz and Capozziello, Salvatore},
TITLE = {The Scavenger Hunt for Quasar Samples to Be Used as Cosmological Tools},
JOURNAL = {Galaxies},
VOLUME = {12},
YEAR = {2024},
NUMBER = {1},
ARTICLE-NUMBER = {4},
URL = {https://www.mdpi.com/2075-4434/12/1/4},
ISSN = {2075-4434},
DOI = {10.3390/galaxies12010004}
}

@article{10.1093/mnras/stv1229,
    author = {Dainotti, M. and Petrosian, V. and Willingale, R. and O'Brien, P. and Ostrowski, M. and Nagataki, S.},
    title = {Luminosity--time and luminosity--luminosity correlations for GRB prompt and afterglow plateau emissions},
    journal = {Monthly Notices of the Royal Astronomical Society},
    volume = {451},
    number = {4},
    pages = {3898-3908},
    year = {2015},
    month = {06},
    issn = {0035-8711},
    doi = {10.1093/mnras/stv1229},
    url = {https://doi.org/10.1093/mnras/stv1229},
}

@article{PhysRevD.103.103509,
  title = {Running Hubble tension and a H0 diagnostic},
  author = {Krishnan, C. and \'O Colg\'ain, E. and Sheikh-Jabbari, M. M. and Yang, Tao},
  journal = {Phys. Rev. D},
  volume = {103},
  issue = {10},
  pages = {103509},
  numpages = {8},
  year = {2021},
  month = {May},
  publisher = {American Physical Society},
  doi = {10.1103/PhysRevD.103.103509},
  url = {https://link.aps.org/doi/10.1103/PhysRevD.103.103509}
}

@misc{krishnan2022h0universalflrwdiagnostic,
      title={$H_0$ as a Universal FLRW Diagnostic}, 
      author={Chethan Krishnan and Ranjini Mondol},
      year={2022},
      eprint={2201.13384},
      archivePrefix={arXiv},
      primaryClass={astro-ph.CO},
      url={https://arxiv.org/abs/2201.13384}, 
}

@ARTICLE{2017A&A...600A..98D,
       author = {{Dainotti}, M.~G. and {Nagataki}, S. and {Maeda}, K. and {Postnikov}, S. and {Pian}, E.},
        title = "{A study of gamma ray bursts with afterglow plateau phases associated with supernovae}",
      journal = {aap},
         year = 2017,
        month = apr,
       volume = {600},
          eid = {A98},
        pages = {A98},
          doi = {10.1051/0004-6361/201628384},
archivePrefix = {arXiv},
       eprint = {1612.02917},
 primaryClass = {astro-ph.HE},
       adsurl = {https://ui.adsabs.harvard.edu/abs/2017A&A...600A..98D}
}

@misc{jia2025hubbletensionresolveddesi,
      title={The Hubble Tension resolved by the DESI Baryon Acoustic Oscillations Measurements}, 
      author={X. D. Jia and J. P. Hu and D. H. Gao and S. X. Yi and F. Y. Wang},
      year={2025},
      eprint={2509.17454},
      archivePrefix={arXiv},
      primaryClass={astro-ph.CO},
      url={https://arxiv.org/abs/2509.17454}, 
}

@article{Brevik_2017,
   title={Viscous cosmology for early- and late-time universe},
   volume={26},
   ISSN={1793-6594},
   url={http://dx.doi.org/10.1142/S0218271817300245},
   DOI={10.1142/s0218271817300245},
   number={14},
   journal={International Journal of Modern Physics D},
   publisher={World Scientific Pub Co Pte Ltd},
   author={Brevik, Iver and Grøn, Øyvind and de Haro, Jaume and Odintsov, Sergei D. and Saridakis, Emmanuel N.},
   year={2017},
   month=dec, pages={1730024} }

@article{Capozziello_2006,
   title={Observational constraints on dark energy with generalized equations of state},
   volume={73},
   ISSN={1550-2368},
   url={http://dx.doi.org/10.1103/PhysRevD.73.043512},
   DOI={10.1103/physrevd.73.043512},
   number={4},
   journal={Physical Review D},
   publisher={American Physical Society (APS)},
   author={Capozziello, S. and Cardone, V. F. and Elizalde, E. and Nojiri, S. and Odintsov, S. D.},
   year={2006},
   month=feb }

@article{Nojiri_2005,
   title={Inhomogeneous equation of state of the universe: Phantom era, future singularity, and crossing the phantom barrier},
   volume={72},
   ISSN={1550-2368},
   url={http://dx.doi.org/10.1103/PhysRevD.72.023003},
   DOI={10.1103/physrevd.72.023003},
   number={2},
   journal={Physical Review D},
   publisher={American Physical Society (APS)},
   author={Nojiri, Shin’ichi and Odintsov, Sergei D.},
   year={2005},
   month=jul }

@article{Brevik_2011,
   title={Viscous little rip cosmology},
   volume={84},
   ISSN={1550-2368},
   url={http://dx.doi.org/10.1103/PhysRevD.84.103508},
   DOI={10.1103/physrevd.84.103508},
   number={10},
   journal={Physical Review D},
   publisher={American Physical Society (APS)},
   author={Brevik, I. and Elizalde, E. and Nojiri, S. and Odintsov, S. D.},
   year={2011},
   month=nov }

@ARTICLE{1998GrCo....4S..50D,
       author = {{Dymnikova}, I. and {Khlopov}, M. Yu.},
        title = "{Self-consistent initial conditions in inflationary cosmology.}",
      journal = {Gravitation and Cosmology},
         year = 1998,
        month = jan,
       volume = {4},
        pages = {50-55},
       adsurl = {https://ui.adsabs.harvard.edu/abs/1998GrCo....4S..50D}
}

@article{DYMNIKOVA_2000,
   title={DECAY OF COSMOLOGICAL CONSTANT AS BOSE CONDENSATE EVAPORATION},
   volume={15},
   ISSN={1793-6632},
   url={http://dx.doi.org/10.1142/S0217732300002966},
   DOI={10.1142/s0217732300002966},
   number={38n39},
   journal={Modern Physics Letters A},
   publisher={World Scientific Pub Co Pte Lt},
   author={Dymnikova, Irina and Khlopov, Maxim},
   year={2000},
   month=dec, pages={2305-2314} }

@article{Dymnikova:2001jy,
    author = "Dymnikova, I. and Khlopov, M.",
    title = "{Decay of cosmological constant in selfconsistent inflation}",
    doi = "10.1007/s100520100625",
    journal = "Eur. Phys. J. C",
    volume = "20",
    pages = "139--146",
    year = "2001"
}

@article{10.1093/mnras/211.2.277,
    author = {Doroshkevich, A. G. and Khlopov, M. Yu.},
    title = {Formation of structure in a universe with unstable neutrinos},
    journal = {Monthly Notices of the Royal Astronomical Society},
    volume = {211},
    number = {2},
    pages = {277-282},
    year = {1984},
    month = {11},
    issn = {0035-8711},
    doi = {10.1093/mnras/211.2.277},
    url = {https://doi.org/10.1093/mnras/211.2.277},
}

@ARTICLE{1985SvAL...11..236D,
       author = {{Doroshkevich}, A.~G. and {Khlopov}, M.~Y.},
        title = "{Fluctuations of the Cosmic Background Temperature in Unstable-Particle Cosmologies}",
      journal = {Soviet Astronomy Letters},
         year = 1985,
        month = aug,
       volume = {11},
        pages = {236-238},
       adsurl = {https://ui.adsabs.harvard.edu/abs/1985SvAL...11..236D}
}

@ARTICLE{1988SvA....32..127D,
       author = {{Doroshkevich}, A.~G. and {Klypin}, A.~A. and {Khlopov}, M.~Y.},
        title = "{Cosmological Models with Unstable Neutrinos}",
      journal = {\sovast},
         year = 1988,
        month = apr,
       volume = {32},
        pages = {127},
       adsurl = {https://ui.adsabs.harvard.edu/abs/1988SvA....32..127D}
}

@article{Riess:2016jrr,
    author = "Riess, Adam G. and others",
    title = "{A 2.4{\%} Determination of the Local Value of the Hubble Constant}",
    eprint = "1604.01424",
    archivePrefix = "arXiv",
    primaryClass = "astro-ph.CO",
    doi = "10.3847/0004-637X/826/1/56",
    journal = "Astrophys. J.",
    volume = "826",
    number = "1",
    pages = "56",
    year = "2016"
}

@article{Efstathiou:2021ocp,
    author = "Efstathiou, George",
    title = "{To H0 or not to H0?}",
    eprint = "2103.08723",
    archivePrefix = "arXiv",
    primaryClass = "astro-ph.CO",
    doi = "10.1093/mnras/stab1588",
    journal = "Mon. Not. Roy. Astron. Soc.",
    volume = "505",
    number = "3",
    pages = "3866--3872",
    year = "2021"
}

@article{Visser:2003vq,
    author = "Visser, Matt",
    title = "{Jerk and the cosmological equation of state}",
    eprint = "gr-qc/0309109",
    archivePrefix = "arXiv",
    doi = "10.1088/0264-9381/21/11/006",
    journal = "Class. Quant. Grav.",
    volume = "21",
    pages = "2603--2616",
    year = "2004"
}

@article{Visser:2004bf,
    author = "Visser, Matt",
    editor = "McClelland, D. E. and Scott, S. M.",
    title = "{Cosmography: Cosmology without the Einstein equations}",
    eprint = "gr-qc/0411131",
    archivePrefix = "arXiv",
    doi = "10.1007/s10714-005-0134-8",
    journal = "Gen. Rel. Grav.",
    volume = "37",
    pages = "1541--1548",
    year = "2005"
}
\end{document}